\documentclass[aip,jcp,reprint,showpacs,amsmath,amssymb]{revtex4-1}

\usepackage{graphicx}
\usepackage{dcolumn}
\usepackage{bm}
\usepackage{datetime}
\usepackage[usenames,dvipsnames,svgnames,table]{xcolor}
\usepackage{booktabs}

\usepackage[breaklinks,colorlinks,linkcolor=black,citecolor=blue,urlcolor=blue]{hyperref}

\newcommand*{\plimsoll}{{\ensuremath{-\kern-4pt{\ominus}\kern-4pt-}}}

\begin{document}

\preprint{AIP/123-QED}

\title{Multi-scale coarse-graining for the study of assembly pathways in DNA-brick self-assembly}

\author{Pedro Fonseca}
\email{pedro.fonseca@physics.ox.ac.uk}
\affiliation{Rudolf Peierls Centre for Theoretical Physics, University of Oxford, 1 Keble Road, Oxford, OX1 3NP, United Kingdom}

\author{Flavio Romano}%
\affiliation{ Dipartimento di Scienze Molecolari e Nanosistemi,
Universit{\` a} Ca' Foscari, Via Torino 155, 30172 Venezia Mestre, Italy.}

\author{John S. Schreck}%
\affiliation{Department of Chemical Engineering, Columbia University, 500 W 120th St, New York, NY 10027, USA}

\author{Thomas E. Ouldridge}%
\affiliation{ Department of Bioengineering, Imperial College London, London, SW7 2AZ, United Kingdom}

\author{Jonathan P. K. Doye}
\affiliation{Physical and Theoretical Chemistry Laboratory, Department of Chemistry, University of Oxford, South Parks Road, Oxford, OX1 3QZ, United Kingdom}

\author{Ard A. Louis}
\affiliation{Rudolf Peierls Centre for Theoretical Physics, University of Oxford, 1 Keble Road, Oxford, OX1 3NP, United Kingdom}


\date{\today}

\begin{abstract}

Inspired by recent successes  using single-stranded DNA tiles to produce complex structures,  we develop a two-step coarse-graining approach that uses detailed thermodynamic calculations with oxDNA, a nucleotide-based model of DNA, to parametrize a coarser kinetic model that can reach the time and length scales needed to study the assembly mechanisms of these structures. We test the model by performing a detailed study of the assembly pathways for a two-dimensional target structure made up of 334 unique strands each of which are 42 nucleotides long. Without adjustable parameters, the model reproduces a critical temperature for the formation of the assembly that is close to the temperature at which assembly first occurs in experiments.   Furthermore, the model allows us to investigate in detail the nucleation barriers and the distribution of critical nucleus shapes for the assembly of a single target structure. The assembly intermediates are compact and highly connected (although not maximally so) and classical nucleation theory provides a good fit to the height and shape of the nucleation barrier at temperatures close to where assembly first occurs. 

\end{abstract}

\maketitle

\section{\label{sec:level1}Introduction}

Multi-component self-assembly has emerged over the past years as a promising route towards the fabrication of supramolecular structures at the nanoscale. Unlike conventional crystals and many self-assembling systems in soft matter physics where a small number of building block types is used many times to assemble a desired product, in such \textit{addressable assembly},\cite{Cadamartiri15,Frenkel2014,Jacobs_review_paper} the use of unique building blocks  (programmed to occupy a unique ``address" in the desired target) offers direct control over shape, size and functionality, thus allowing for a much wider design space. 

One of the most popular ways to achieve addressable multi-component assembly is with short strands of DNA, by virtue of the specificity of its interactions.\cite{Wang2016,Rothemund_origami,PengYin2d} Typically, each DNA strand  is divided into various domains, each programmed (through the choice of sequence) to hybridize with a different neighbouring strand in the target structure. By designing which of the strands hybridize with each of the other strands, the shape and composition of the target structure can be uniquely defined. Besides offering the necessary specificity to codify interactions, the use of DNA also ensures that bonds between building blocks also have a degree of reversibility, granting self-assembling systems some capacity for self-correction if undesirable bonds eventually form.

DNA origami,\cite{Rothemund_origami} where many short strands are designed to bind to a long single-stranded scaffold, has been the most used technique for designing large self-assembling structures in DNA nanotechnology.\cite{Rothemund_origami,origami_review} More recently, a new scaffold-free method that uses only short DNA strands has been developed by the Yin group. The single-stranded tile (or DNA brick) approach involves assembling large three dimensional structures with  intricate shapes and  complex functionalities from  thousands of unique strands, using a one-pot thermal annealing protocol.\cite{PengYin2d,PengYin3d,Peng_Yin_Dephts,Peng_yin_10nm,Peng_yin_tubes,Peng_yin_isothermal,Peng_yin_design_space,Peng_yin_reconfigurations,Peng_yin_10000}  

 This addressable self-assembly method, may, at first glance, seem straightforward, since each unit is designed to be in just one unique position. In common with other examples of self assembly of finite objects with highly specified interactions,\cite{Parent2005,Wilber_2009,Hagan_2014aa,perlmutter2015mechanisms} the relative rates of nucleation and growth need to be carefully controlled. The nucleation barrier needs to be low enough so that it can be crossed on experimentally accessible time scales, whilst also being large enough that the growth time for structures to reach their equilibrium shape is significantly shorter than the time for the next nucleation event. If the latter does not hold, the system will rapidly grow many more partially formed nuclei than the maximum number of possible assembled structures, thus using up available monomers (monomer starvation). 

However, there are also additional hurdles to correct assembly that are specific to addressable assembly. Firstly, the extra entropy associated with the large number of component strands pushes the assembly transition down to lower temperatures compared to an equivalent one-component system.
Secondly, although incorrect binding is a potential problem for any kind of self-assembly, the number of pairs of components between which misbonding could occur is vastly increased.  Thus, the  strength of the correct interactions must be strong enough in comparison to undesirable interactions to compensate for the vastly larger number of potential incorrect interactions. Although DNA provides an ideal material to have the requisite specificity of interactions, it is not inherently obvious that there will be an assembly window where the conditions that there is sufficient thermodynamic driving force and yet misbonding is not prevalent are both satisfied. Thus, the success of the first demonstrations of the single-stranded tile (SST) method was conceptually somewhat surprising.
 
The need to explain the success of the SST assembly method inspired a range of important theoretical work.\cite{Jacobs_review_paper} For example, Reinhardt and Frenkel, using a lattice model\cite{lattice_model1} where individual strands of DNA are modelled as patchy particles with rigid patches representing binding domains, were first to demonstrate  the successful assembly of a large structure in a model of addressable assembly. This same model was later used to explore the effect of the coordination number on the nucleation barriers, and hence the assembly success.\cite{lattice_model2} An off-lattice model,\cite{Aleks_off_lattice} which provides a somewhat more robust description of the translational and rotational entropy of the single strands, was also found to reproduce the main physical phenomena observed in the simpler lattice model.   
 
In parallel to these numerical investigations, theoretical calculations by Jacobs \textit{et al.},\cite{lattice_model_wang2} helped explain why a time-dependent annealing protocol is necessary to achieve a fully assembled structure.  Due to the large combinatorial entropy for these systems, as temperature is cooled and the critical nucleation barrier is crossed, the free-energy minimum is for a partially formed structure. Further cooling is needed for the final structure to completely form.  Thus addressable assembly differs from more conventional assembly with a small number of sub-units, where complete assembly can typically occur at a single temperature. Similar theoretical calculations\cite{lattice_model_wang} were also used to predict how nucleation barriers depend on the topology of the graph describing how  the particles are connected in the assembled state. More recently, this model was used  to investigate strategies to optimise nucleation and growth of unbounded, periodic structures.\cite{Jacobs_protocol_design}  Another strand of research into addressable assembly, also using simplified models,\cite{Murugan_undesired_usage} has suggested that using carefully chosen non-stoichiometric abundances of assembly components may help to mitigate aggregation and undesirable monomer starvation effects. 

 While these calculations have provided important qualitative insights, these minimal models neglect important aspects of these systems, such as the polymeric nature of the constituents, thus preventing quantitative predictions.  More detailed modelling would be beneficial to help guide SST experiments and to work out  the relative importance and the interplay of the various mechanisms that affect the addressable assembly of a large structure from a set of individual single strands of DNA.

\begin{figure}[t]
    \centering
    \includegraphics[width=0.45\textwidth]{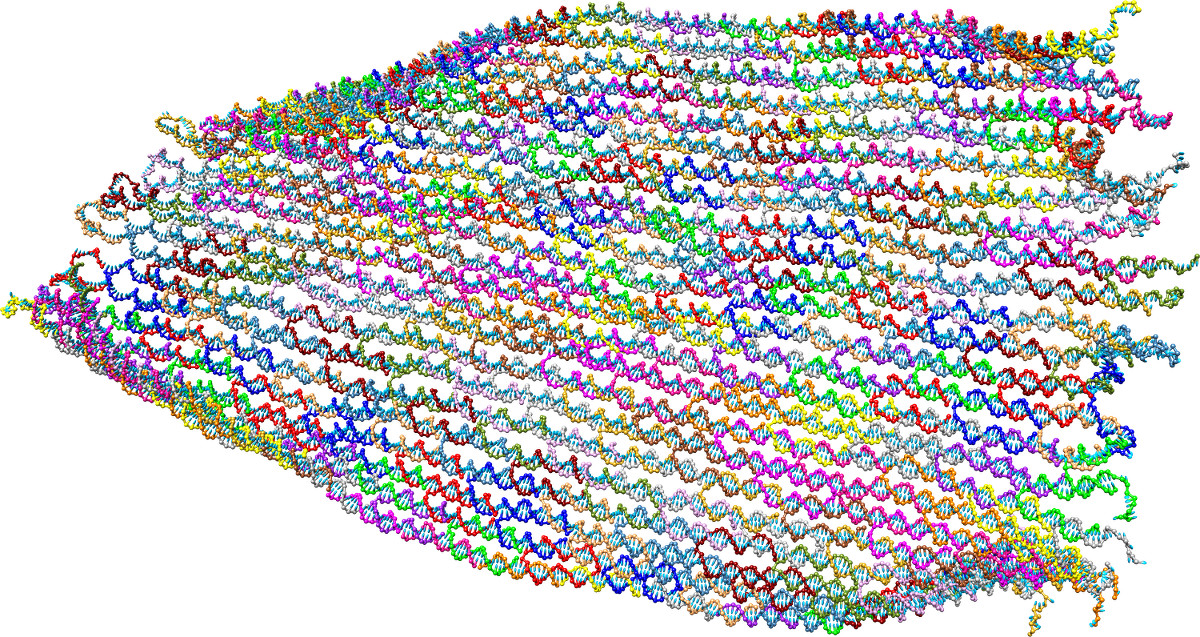}
    \caption{A typical oxDNA configuration illustrating the fully assembled molecular canvas at $T=25^\circ \textnormal{C}$.  This snapshot was taken after allowing an initially fully planar configuration to equilibrate using molecular dynamics simulations with oxDNA2,\cite{snodin2015introducing} an extension of the original oxDNA that allows for a better description of structural properties of large (kilobase-pair) structures. Distinct tile species are depicted using different colours. At this low temperature the entire 334 strand structure is highly prevalent in equilibrium under typical conditions.}\label{fig:membrane}
\end{figure}

The fundamental reaction exploited by the SST assembly method is the hybridization of different domains between separate single strands. It is natural to model these reactions using oxDNA,\cite{tom_model_jcp,oxdna_review} a nucleotide-level model of DNA that has successfully been applied to a diverse set of systems and processes where hybridization plays a key role, ranging from DNA devices such as nanotweezers,\cite{ouldridge2010dna} and DNA walkers\cite{ouldridge2013optimizing,sulc2014simulating} to the dynamics of the displacement reaction\cite{srinivas2013biophysics,Machinek:2014aa} to  the breaking of duplex DNA under force\cite{romano2013coarse} to hybridization without\cite{ouldridge2013dna} and with\cite{schreck2015dna} secondary structure,  and even to the assembly of small origamis.\cite{snodin2016direct} This success, and the good agreement (where comparison is possible) with experiment, suggests that oxDNA can be used to understand SST assembly.      

On the other hand, the scale of the challenge to model self-assembly for these systems is evident. For example, the two-dimensional SST rectangle, first presented in Ref.~\onlinecite{PengYin2d} and illustrated in Fig. 1 using oxDNA, is made up of 334 unique single strands each 42 nucleotides long giving 14,028 nucleotides in total. Each strand is composed of four  concatenated domains of length 10-11-11-10 bases or 11-10-10-11 bases (depending if the strand occupies an odd or even row) that bond with four local neighbours during self assembly.      This rectangle serves as a molecular canvas, from which custom shapes can be created by removing certain strand types.  Using this strategy, over one hundred different shapes were successfully assembled using  a one-pot annealing protocol where the temperature is lowered at a fixed rate.\cite{PengYin2d}  While computer simulations with oxDNA can be used to study structural properties of a fully assembled rectangle, as illustrated in Fig. ~\ref{fig:membrane}, it is computationally unfeasible to characterise the spontaneous self-assembly of this structure from a bath of single strands via direct simulations, as the correspondent time scales are far too long even given the speedups that coarse-graining at the level of oxDNA provides. For example, the fastest experimental protocols that lead to correct assembly are on the order of $1^{\circ}\textnormal{C}$ per hour\cite{SobczakDNAfolding}, while oxDNA time steps are on the order of several fs.

To make progress, further coarse-graining  is necessary.  We are assisted by the observation that the equilibrium ensemble of two complementary DNA strands at ideal conditions is dominated by states in which either most of the base pairs are formed between complementary strands, or no base pairs are formed, separated by a free-energy barrier that emerges mostly due to the sudden reduction of translational freedom when the first base pairs form. This feature of DNA hybridization is well illustrated by the successes of a two-state model\cite{SantaLucia1998} in predicting the thermodynamic properties of DNA duplexes and other DNA motifs.

 This time-scale separation suggests that a kinetic model at the level of the chemical master equation, with each state specifying a bonding configuration, may capture the dominant physics of assembly. We are  inspired here by recent success of a kinetic model for DNA origami,\cite{Origami_plus_model,dannenberg2015modelling} where excellent agreement with experiment was achieved.

Our basic coarse-graining strategy for SST assembly of a 2D canvas is to first estimate the rates for key reactions using oxDNA, and then use these rates to parameterise a coarser kinetic model that allows us to work out the full dynamics of assembly.  The ratio of the on to off rates can be estimated using  the free energy of hybridization together with a two-state approximation for the kinetics. While the connectivity network of assembly units seems to be in the natural level of description from both kinetic and structural perspectives, a problem arises because the number of possible non-equivalent connectivity networks grows exponentially with the number of subunits composing the target structure.

To circumvent this problem of determining transition rates between every pair of configurations we identify a finite set of local changes in the connectivity network associated with the binding/unbinding of a staple, and parameterise all transition rates according to which member of this finite set they correspond.  This approach means that  we do not need all theoretically possible transition rates, but instead can treat self-assembly using a limited database of tile association/dissociation rates.   We then use  standard kinetic Monte Carlo (KMC) techniques to sample over all possible states.
 
\begin{figure*}[t]
    \centering
    \includegraphics[width=1.00\textwidth]{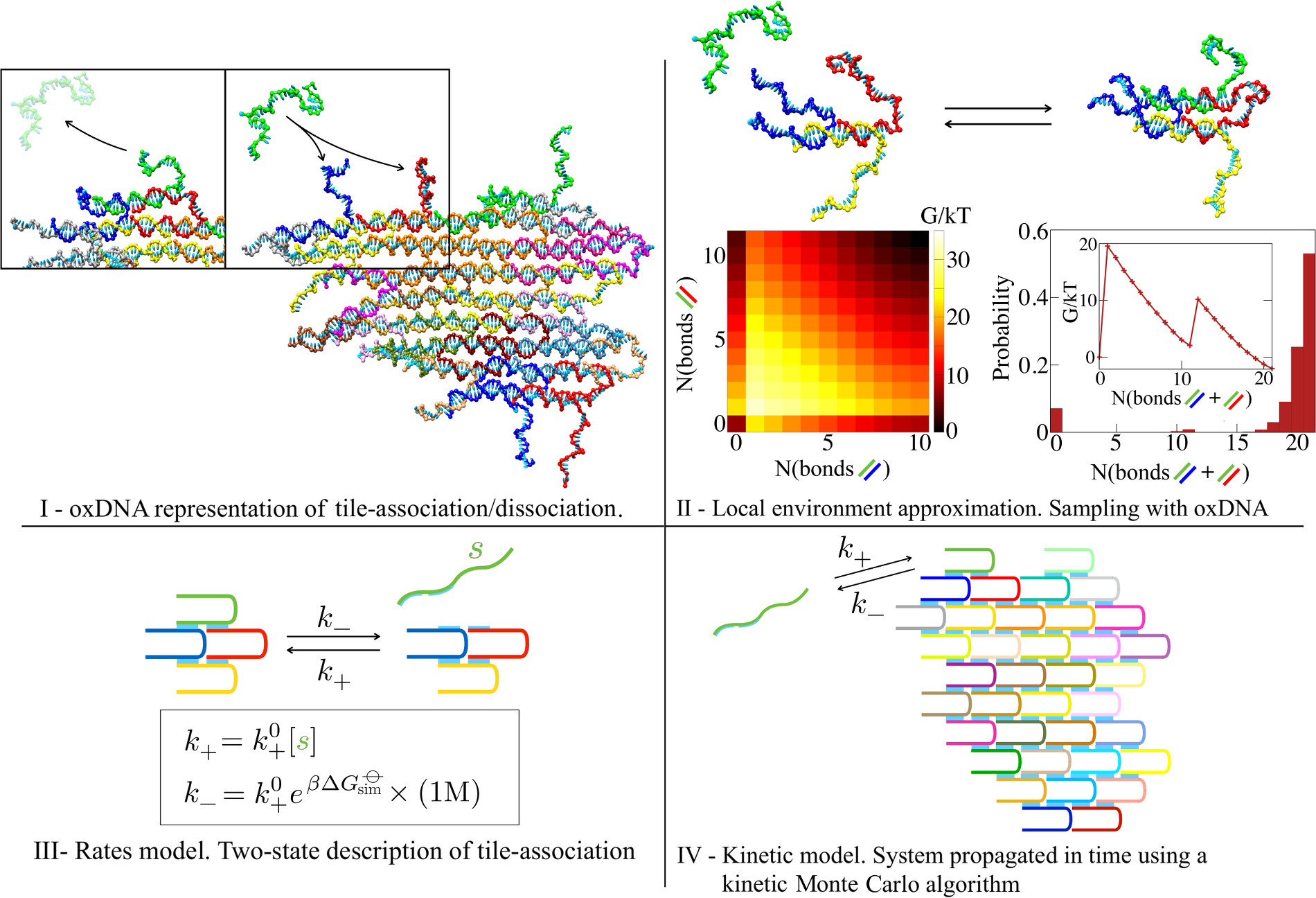}
    \caption{ {\bf Multi-scale coarse-graining approach for tile-based self-assembly}.  \textbf{I,II - top)} The free energy of the tile-association/dissociation reaction is determined by considering the minimal assembly fragment that contains the first neighbours of the assembly site to which the tile associates/dissociates. In this example, the site occupied by the (binding) green strand has two first neighbours (the blue and red strands). The yellow strand (II - top) is also used in order to keep both strands arranged correctly.  We use oxDNA to determine the thermodynamics of this reaction: \textbf{II - bottom left)} The free-energy landscape for tile association as a function of the number of bonded base pairs between complementary domains at $T = 50^\circ \text{C}$, and for the strand concentration used in experiments ($100\textnormal{nM}$).  \textbf{II - bottom right)} Probability profile at equilibrium as a function of the total number of base pairs between complementary domains at $T = 50^\circ \text{C}$. From the plot: $P(N=0)\cong 0.07$, $P(1\leq N\leq 11)\cong 0.01$, $P(12\geq N)\cong 0.92$. Inset: the free-energy profile as a function of the total number of base pairs between complementary domains at $T = 50^\circ\text{C}$. \textbf{III)} The statistics of tile association derived using oxDNA are used to parametrize a two-state description, where a tile is considered to be either ``bound'' or ``diffusing". We model rates by assuming that tiles bind to a cluster, if a complementary domain is present, at rate $k_{+}^0[s]$ where $[s]$ is the concentration of strand-species $s$ and $k_{+}^0$ is assumed to be independent of binding site and temperature, and use the free-energy change of reaction to determine the respective dissociation rates.
\textbf{IV )} Having developed a database of tile-association/dissociation rates, we use a kinetic Monte-Carlo algorithm to propagate the system in time. }\label{fig:cg_scheme}
\end{figure*}

The full  multi-scale procedure to coarse grain our kinetic model and to calculate the dynamics of assembly is summarised in Fig.~\ref{fig:cg_scheme}. The resulting KMC trajectories can easily reach the time-scales of multiple hours used in the assembly experiments\cite{PengYin2d,SobczakDNAfolding} in just a few minutes of computer time.  The only free parameter is the on-rate $k_+^0$  which we assume to be temperature independent, and set by recent experiments of DNA strand displacement reactions.\cite{winfree_kinetics}  This parameter sets time scales, but not free energies (and therefore transition temperatures).    Our calculations, which otherwise have no adjustable parameters, predict a critical temperature that is only a few degrees Celsius different from that measured in experiment.

We proceed as follows.  In the methods section, we describe our coarse-graining methods in more detail, explaining how we define the key states of our model, how we calculate  free-energy differences using oxDNA, how we incorporate these thermodynamic calculations into  our kinetic rates model, and how we use enhanced sampling KMC methods. We also discuss  the importance of some effects we neglect in our multi-scale coarse-graining procedure, including a discussion of the role of misbonding.  In our results section, we first consider the assembly (and disassembly) of a single target structure during both temperature ramp and isothermal protocols. Next we consider the nucleation pathway in more detail, calculating critical nucleus sizes and free-energy profiles as a function of assembly size for different temperatures.   Finally, we present some conclusions, pointing to future applications of our new method.

\section{\label{sec:level1}Methods}

\subsection{\label{sec:level2}Definition of the states}

In our simulations, we follow the assembly and disassembly of a single structure through monomer association from, and dissociation into, a large bath of monomers. We  define the state of the system as a single vector made up of $N$ components $\vec{x} =(x_{1},x_{2},\cdots ,x_{N})$, where $x_{i}=1$ if strand species $i$ is in  the assembly and $x_{i}=0$ otherwise, and $N=334$ is the number of 42-base-long strands in the fully assembled 2D structure that we are studying here. We can make the simplification that the states only describe strands either  fully on or off because of  the essential two-state nature of strand association: the subensemble of states in which programmed hydrogen bonds are formed between the incoming strand and its binding domains in the assembly is statistically dominated by states in which all complementary domains are hybridized and most of the base pairs are formed (see details in Section II.B.2).

The transition rates are denoted by $k(\vec{x}|\vec{y})$ for transitions $\vec{y}\rightarrow \vec{x}$. Possible transitions in the model are those in which a single strand, from now on also called a monomer, either binds to or is removed from the assembly. For consistency, however, transitions in which the assembly is divided into two disconnected parts due to the removal of a monomer are not allowed as we do not describe the reverse process. 

The kinetic simulations performed throughout this work were initiated (unless stated otherwise) from the state $x_{\mu}=\delta_{\mu \nu}$, where $\nu$ was selected randomly from the set of 334 different tile species $(\nu \in [1,334])$, that corresponds to a single tile. After the initial step of the simulation is performed (the association of a second tile), the first tile is allowed to dissociate,  but we always keep at least one tile in the state vector.  Typically a great number of association and dissociation steps occur before any nucleation events happen, so that memory of  the identity of the initial tile is effectively erased.

By using the state definition described above, we focus our attention on the assembly of a single target structure,  while the concentration of strand species is kept constant. Our model is therefore appropriate to describe the first stages of an experimental assembly setup, when the first structures start to form and there is little variation in the concentration of monomers. To capture the full complexity of an assembly process, future work needs to describe explicitly the variation with time of the composition of the assembly mixture, which contains multiple assembly fragments and (usually) a fixed number of monomers. In the present work, we instead focus our attention on the nature of the assembly pathways for single-target assembly, and leave consideration of the collective aspects of the assembly process for future work.

\subsection{\label{sec:level2}Transition rate model}

\subsubsection{\label{sec:level3}Tile association rates}
The state space of the model is based on the assumption that the rearrangement of a  strand to form secondary domains is fast compared to the initial binding of the first domain when a strand arrives from the dilute, well-mixed solution. Furthermore, association rates of duplexes are known to be less sensitive to sequence and conditions than dissociation rates.\cite{Zhang_kinetics, tom_inc_1,tom_nm_1,schreck2015dna}  In the spirit of minimal parameterization of the model, therefore, we assume that tile-association rates do not vary with temperature and use the standard value for the second order rate constant of $k_+^0=6\times 10^{6}\textnormal{M}^{-1}\textnormal{s}^{-1}$, that yields the best fit\cite{winfree_kinetics} to a set of experimental data on second-order hybridization reactions that are part of a displacement reaction system (see details in Section S.X).  Therefore, for a state $\vec{y}$ that is obtained from $\vec{x}$ by adding a single tile, 
we estimate the tile-association rate as
\begin{equation}
k(\vec{y}|\vec{x}) = k_+=k_+^0 [{s}],
\end{equation}where $[{s}]=100\hspace{1mm}\textnormal{nM}$ is the concentration of strand species $s$ (assumed the same for all strand species) in the systems studied here.

\subsubsection{\label{sec:level3}Tile dissociation rates}

If tile association is treated as simple bimolecular binding reaction, as here, then dissociation rates for the reaction $\vec{y} \rightarrow \vec{x}$ can be related to association rates via
\begin{equation}
\frac{k(\vec{x}|\vec{y})}{k(\vec{y}|\vec{x})}= \frac{k_-(\vec{x},\vec{y})}{k_+^0[s]} = e^{\beta \Delta G^\plimsoll(\vec{x},\vec{y})} \times \frac{1M}{[s]}.
\label{eq:deltaG}
\end{equation}
Here $  \Delta G^\plimsoll(\vec{x},\vec{y})$ is the standard free-energy change of formation for the reaction $\vec{x} \rightarrow \vec{y}$, and ${k_-(\vec{x},\vec{y})}$ is the rate of dissociation of a strand from $\vec{y}$ to form $\vec{x}$.  This quantity takes into account the balance between the entropic cost of bringing a diffusing strand into contact, in the correct orientation, and the enthalpic gain due to the formation of hydrogen bonds and stacking interactions between adjacent bases upon tile association.   

\begin{figure}[t]
    \centering
    \includegraphics[width=0.45\textwidth]{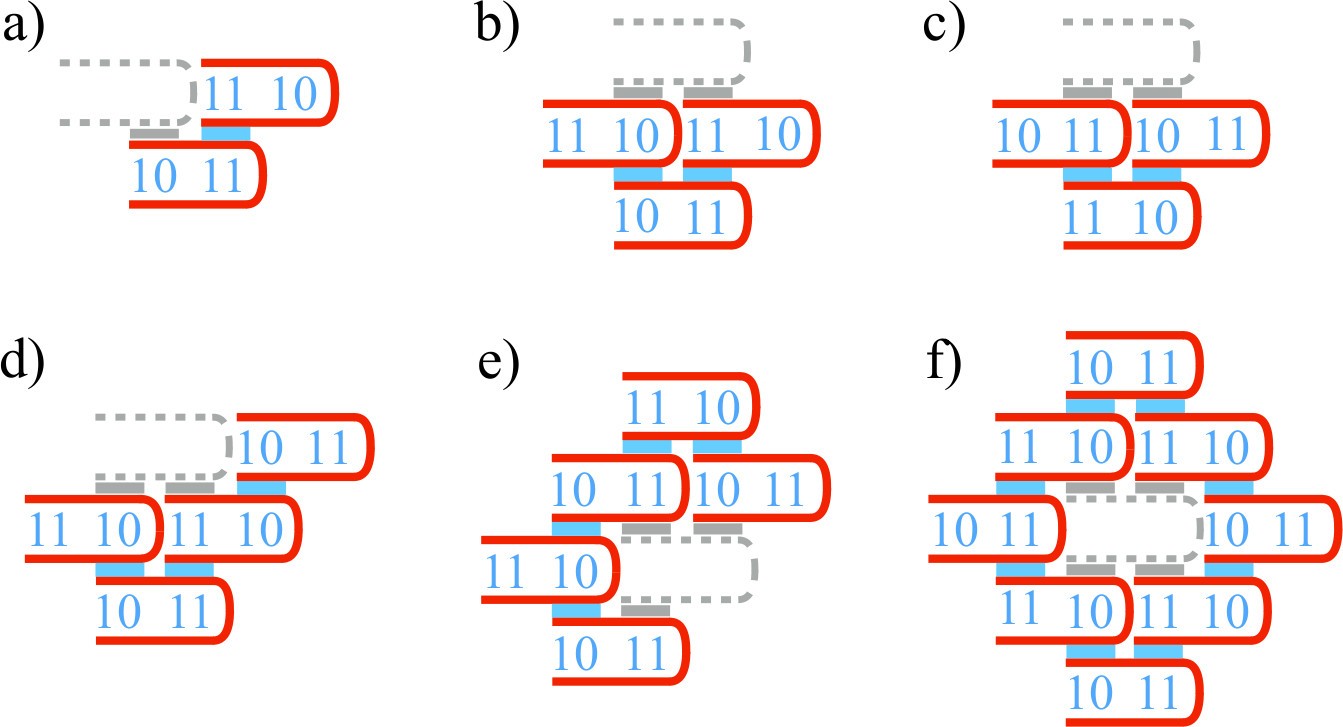}
    \caption{Examples of local environments for tile association.  The red tiles denote the assembly fragment that contains the first neighbours of the binding site (coloured grey). The sections in the assembly fragment that are hybridized are shown in blue. The numbers 10 and 11 indicate the length (in bases) of the binding domains positioned above. The examples involve (a) one, (b)-(d) two, (e) three and (f) four binding domains. The full list of 32 local environments used can be found in the supplementary Section S.VI.}\label{fig:le_examples}
\end{figure}

The major biophysical input into our model is the use of oxDNA simulations to estimate $ \Delta G^\plimsoll(\vec{x},\vec{y})$, and hence ${k_-(\vec{x},\vec{y})}$, for each tile dissociation reaction. While oxDNA can be used to directly calculate  relative rates using either direct molecular dynamics calculations or using techniques such as forward flux sampling (FFS),\cite{hairpin_stacking,Machinek:2014aa,schreck2015dna,srinivas2013biophysics,design_knotted_DNA} for the  relatively simple reaction rates that we need to calculate, an equilibrium calculation of free-energy difference in Eq.~\ref{eq:deltaG}, combined with an estimate of hybridization rates from the literature, is more practical.

In principle oxDNA could be  directly used for all possible tile-association events, but since the assembly can proceed in a very large number of different ways there are far too many possible transitions to make this direct approach feasible. Additionally, for larger tile-fragments, a lot of the computational load would involve degrees of freedom that are far from the site where hybridization is occurring.   To simplify our calculations, we apply a \textit{local environment approximation} in which  we assume that the enthalpic and entropic factors that govern $\Delta G^\plimsoll(\vec{x},\vec{y})$ can be approximated by considering the minimal assembly fragment that contains the first neighbours of the attaching site in the assembly (see Figs.~\ref{fig:cg_scheme}.I and \ref{fig:cg_scheme}.II).

 Using this approximation, each possible transition in the model falls within a relatively small number of distinct cases (local environments), each specified by the number, size, and position of binding domains that hybridize during tile association, and the number and position of bases adjacent to the binding domains on the assembly that coaxially stack. 
 Several examples of local environments are illustrated in Fig.~\ref{fig:le_examples}, and the full set  of 32 local geometries that we use is given in the supplementary information (Section S.VI).

For each local environment, we determine the free energy of tile association by simulating the minimal assembly fragment that contains all first neighbours of the binding site, and the incoming tile with the correct sequence, in a periodic box using oxDNA. In the simulations we use the average-base version of the model,\cite{tom_model_jcp} which allows complementary base-pairing between AT and CG bases, but averages over the hydrogen-bonding and stacking parameters. Although tile-binding energies are expected to depend on the tile's sequence, we use this approximation in order to focus our attention on generic mechanisms of tile association rather than the effect of using a particular sequence. In addition, the simulations are set up so that the formation of secondary structure for single strands is prevented.  The advantage of these simplifications is that only one calculation is needed for each local geometry.  Sequence dependence can easily be included into oxDNA,\cite{vsulc2012sequence} but for simplicity we have not done this here.   Future work may include sequence dependence. This may shine light on conflicting predictions in the literature, with some arguing that adding sequence heterogeneity stabilises assembly intermediates,\cite{lattice_model_wang} while others predict that it hinders assembly.\cite{Brenner_SA}

The free-energy profiles were first sampled as a function of the number of native base pairs  between fragment and incoming strand, using the virtual-move Monte Carlo (VMMC) algorithm of Whitelam and Geissler\cite{Whitelam_VMMC_2} that attempts moves of dynamically selected clusters of nucleotides, greatly accelerating sampling in oxDNA over Metropolis Monte Carlo (see details in Section S.II.A). In addition, umbrella sampling \cite{Torrie_Umbrella} was used in order to improve the sampling of rare transitions between free-energy minima (see details in Section S.II.B). In the simulations, a base pair is considered bonded if the respective hydrogen bonding energy falls bellow $-0.596\hspace{1mm} \textnormal{kcal}\hspace{0.7mm}\textnormal{mol}^{-1}$ (which corresponds to about 15\% of the typical hydrogen bond energy). The number of base pairs between designed complementary domains is defined as the order parameter and used to divide the configurational space into regions biased by the umbrella potential. Base pairs that were not designed to be part of the assembled structure are not allowed to form in the simulations. Due to the complexity of reactions in which a tile associates to the assembly through more than one binding domain, the order parameter was augmented with other variables, that specify the distance between the centre of mass of each binding domain of the tile and the respective complementary sequence in the assembly fragment. To improve the efficiency of the calculations, the order parameter space was split into sampling windows, and the statistics were later combined using the weighted histogram analysis method (WHAM).\cite{kumar_wham} The simulations were performed at $T=50^\circ \textnormal{C}$,  which is within the range where the transition temperature was expected to be, and the statistics were extrapolated to other temperatures using single histogram reweighting~\cite{vsulc2012sequence} (see details in Section S.II.C).

For each local environment, the tile-association free energy $\Delta G_{\textnormal{sim}}$  is defined as the free-energy difference between the subensemble of states in which at least one base pair is formed between complementary domains $(\Omega _c)$, and the subensemble of diffusive states, in which no base pairs are formed between the assembly and tile $(\Omega _d)$: $\beta \Delta G_{\textnormal{sim}}=-\textnormal{ln}[p(\Omega_c)/p(\Omega_d)]$. As an example, for the local environment depicted in Fig.~\ref{fig:cg_scheme}, $\Omega _c$ corresponds to the subensemble of states satisfying $N_{10}+N_{11}>0$ where $N_{10}$ and $N_{11}$ denote the number of base pairs formed between complementary domains that are 10  (between green and blue tiles) and 11 (between green and red tiles) nucleotides long, respectively, and $\Omega _d$ corresponds to the subensemble of states satisfying $N_{10}+N_{11}=0$. To explore the accuracy of the two-state approximation, we also consider free-energy landscapes in which the bound state is further subdivided into states with different degrees of bonding. 

The simulations were performed in a small volume $v$, implying a higher effective concentration $1/v$ than in experiments, so as to reduce diffusion times. We made sure that the systems were large enough so that periodic images could not interact.  Afterwards, the tile-association free energies and free-energy profiles were extrapolated to concentrations relevant to experiment by adding the factor $\Delta \Delta G_{\mathrm {vol}}=k_{B}T \ln(u/u_0)$ to the free energies of states with base pairs. Here, $u_0$ is the desired tile concentration and $u$ the tile concentration used in the simulations, which varied between $4.38\times 10^{-6}\textnormal{M}$ (for the smallest systems), and $1.72\times 10^{-7}\textnormal{M}$ (for the largest ones). This extrapolation provides the best description for the thermodynamics of the binding process under the relevant conditions. Finally, we determined the effective tile-association free energies at standard concentration ($\Delta G^\plimsoll$), as employed in the kinetic model. To do so, we employed the method in Ref.~\onlinecite{bulk_extrapolation} that relates bonding probabilities observed in a high concentration, single-replica simulation to $\Delta G^\plimsoll$.

In Fig.~\ref{fig:cg_scheme}.II we show  a typical equilibrium free-energy profile of tile association obtained from the simulations, determined  here for local environment 7 (see Section S.VI). This same environment is used throughout  Fig.~\ref{fig:cg_scheme}. We also present the respective probability profile, projected along a single reaction coordinate that specifies the total number of base pairs formed at $50^\circ \text{C}$. The probability profile in Fig.~\ref{fig:cg_scheme}.II helps illustrate  that the ensemble is statistically dominated by states in which either most base pairs are formed (and all complementary domains hybridized) or no base pairs are formed. This feature is  most pronounced in cases in which the tile binds with the assembly fragment through 3 and 4 binding domains (see Section S.VI).

Given the simple two-state nature of the transitions, one might ask whether a simpler approach could be used to estimate the free energy of tile association. For example, the nearest-neighbour (NN) model of Santa Lucia,\cite{SantaLucia1998} the predictions of which were used to fit the oxDNA model, provides an accurate description of the thermodynamics of hybridization. This model could be used for binding events where a tile is only able to form a single domain (e.g.\ Fig.~\ref{fig:le_examples}a). However, the NN model is unable to provide any estimate of the free-energy barrier to initiating a second (Fig.~\ref{fig:cg_scheme}.II) or subsequent domains (see Section S.VI). The accurate capturing of these barriers, which arise from the loss of conformational entropy of the strands involved on initiating the formation of these additional binding domains, is the key extra ingredient that using the oxDNA model provides. We find that this barrier is typically around $8 kT$.

Similarly, this is the key limitation of previous developed patchy-particle models for SST assembly\cite{lattice_model1,lattice_model2} that use the NN model to provide the strength of the patch-patch interactions. These models do exhibit a free-energy barrier for the formation of a second domain, but this arises from the loss in entropy of the patchy particle assembly (e.g.\ a patchy particle with only one bond is able to freely rotate about this bond, but this degree of freedom becomes constrained on the formation of a second bond) and not from a realistic description of DNA's polymeric degrees of freedom. In Section III.C, we will highlight some of the effects that our more accurate description of the thermodynamics of multiple domain formation has on the observed assembly pathways.

\subsection{Simulation of the coarse-grained kinetic model}
\subsubsection{Kinetic Monte Carlo algorithm}

Having developed a database of rates, the task  shifts to determining the time evolution of the assembly. For that purpose we implemented a standard KMC algorithm.\cite{gillespie_kmc} Starting at assembly state $\vec{x}$, the algorithm determines the nearest neighbourhood of tiles composing the fragment and binding sites at the interface. For tiles composing the fragment,  the algorithm attributes the relevant dissociation rate from the kinetic database determined in advance. If the fragment is composed by only two tiles, a single dissociation rate is attributed that corresponds to the dissociation of the dimer. For sites at the interface the algorithm attributes an association rate (which we assumed to be independent of local environment and temperature). Next, the algorithm determines the cumulative rate function $R(i)=\sum_{j=1}^{i}k(\vec{x_j} |\vec{x})$ of transitions $\vec{x}\rightarrow \vec{x_i}$ allowed, and chooses which transition to carry out by generating a random number $u\in [0,1)$ and finding the transition $i$ that satisfies $R(i-1)< uR(n_{trans}) \leq R(i)$, where $R(n_{trans})$ is the cumulative rate of leaving state $\vec{x}$: $R(n_{trans})=\sum_{i=1}^{n_{trans}}k(\vec{x_i} |\vec{x})$. In the case the  assembly state is a dimer, the rate for dimer break-up is only counted once and the two possible destination states are selected at random if dissociation is picked by the algorithm. Time is then updated, such that $\delta t$ follows the probability distribution for the time of first transition leaving $\vec{x}$: $P(\vec{x},t+\delta t)=\exp ( -R(n_{trans})\delta t )$, which is satisfied by choosing $\delta t=-R(n_{trans})^{-1}\log v$ where $v\in [0,1)$ is a random number. 

\subsubsection{The simulated ensemble}

If the simulation is run for an arbitrarily long time under constant conditions, it reaches a steady state. Here we discuss the nature of that steady state, and its relation to the equilibrium yield of partial assemblies. In an ideal (dilute) solution, in which partial assemblies evolve independently in a large bath of tiles at fixed concentration $[s]$, the steady-state concentration of a cluster $\vec{x}$ is given by the equilibrium condition
 \begin{equation}
{[\vec{x}]} = [s] \left(\frac{[s]}{1{\textrm M}} \right) ^{N_{\vec{x}}-1}   \exp  \left(-\beta G^\plimsoll(\vec{x})\right)
\label{eq:conc}
\end{equation}
in which  $G^\plimsoll(\vec{x})$  is the free energy of assembly of $\vec{x}$ from its constituent strands at standard concentrations, and $N_{\vec{x}} $ is the number of strands in assembly state $\vec{x}$. Perfectly accurate values of $\Delta G^\plimsoll(\vec{x},\vec{y})$ would satisfy
\begin{equation}
\Delta G^\plimsoll(\vec{x},\vec{y}) = G^\plimsoll(\vec{y}) - G^\plimsoll(\vec{x}).
\label{eq:thermoSC}
\end{equation}

In principle, one could study the assembly process given by a well-defined $\Delta G^\plimsoll(\vec{x},\vec{y})$, $k_+^0$ and $[s]$  in a fixed volume, and infer the equilibrium concentration of partial assemblies from the frequency with which they are observed in the simulation volume.\cite{bulk_extrapolation}  Instead of simulating a fixed volume, however,  we pursue an  ``assembly-eye view" of the same process in which we simply track the growth and shrinkage of a single assembly. In this case, the steady-state probability of the system occupying a given state $\vec{x}$ in equilibrium is given by
\begin{equation}
p^{\textnormal{eq}}(\vec{x}) = 
\begin{cases}
\frac{1}{2Z}, \hspace{5mm} N_{\vec{x}} =1\\
\frac{1}{Z} \left(\frac{[s]}{1 \mathrm M}\right)^{N_{\vec{x}}-1} \exp \left( -\beta \Delta G^\plimsoll(\vec{x})\right), \hspace{5mm} N_{\vec{x}} >1,
\end{cases}
\label{eq:ss}
\end{equation}
where $Z$ is a normalising factor. Eq.~\ref{eq:ss} can be verified by observing that, given these probabilities, the net flux of trajectories between all $\vec{x}$ is zero under the simulation procedure (yielding a detailed-balanced steady state). We note that $p^{\textnormal{eq}}(\vec{x})/p^{\textnormal{eq}}(\vec{y})$ is consistent with the equilibrium concentrations in Eq.~\ref{eq:conc} unless either $N_{\vec{x}} =1$ or $N_{\vec{y}}=1$ (in this case, the ratio deviates by a factor of two). In order for the equilibrium probabilities in the ``cluster's-eye view" to map directly into concentrations, we correct, after sampling, the equilibrium probability of finding the system in a state $\vec{x}$ with $N_{\vec{x}}=1$ by a multiplicative factor $2$.

In fact, due to the local approximation to $\Delta G^\plimsoll(\vec{x},\vec{y})$, and the finite sampling accuracy of oxDNA simulations, Eq.~\ref{eq:thermoSC} is only approximately satisfied. Detailed balance is therefore marginally violated in simulations; we perform the following calculation to gauge whether deviations from detailed balance are significant.

Let  $\vec{x}_i$ and $\vec{x}_j$ be two assembly states, and $\wp_{ij}\equiv \vec{x}_i \rightarrow \vec{x}_{i+1} \rightarrow \vec{x}_{i+2} \rightarrow \cdots \rightarrow \vec{x}_{j-1}  \rightarrow \vec{x}_j$ be a reaction pathway that connects $\vec{x}_i$ and $\vec{x}_j$. If detailed balance is satisfied exactly, the total free-energy change 
\begin{equation}
G^\plimsoll(\vec{x}_j) - G^\plimsoll(\vec{x_i}) = \sum_{ \vec{y} \rightarrow \vec{z} \in \wp_{ij}}   \Delta G^\plimsoll(\vec{y},\vec{z}),
\label{eq:G_balance}
\end{equation}
should be pathway-independent. In Section S.VIII we consider a few examples of assembly states $\vec{x}_i$ and $\vec{x}_j$, and determine the factor on the RHS of  Eq.~\ref{eq:G_balance} for different paths $\wp_{ij}$. We find the variation with path to be small for the examples considered (see Section S.VIII). While differences between paths imply that detailed balance is not strictly satisfied, we find these differences to be relatively small and assume that they can be neglected as a first approximation for the remainder of this manuscript. A similar approach was taken in a recent model of DNA origami folding.\cite{Origami_plus_model,dannenberg2015modelling} 

We further gauge the effect of detailed balance deviations by determining free-energy profiles as a function of assembly size using the biasing method introduced in Section II.C.3, which is based upon the assumption of detailed balance, and comparing the resulting profiles with those obtained through unbiased sampling (see Section S.IX). We find good agreement between both methods which further supports our assumption that deviations from detailed balance can be neglected as a good approximation.

\subsubsection{\label{sec:level2}Enhanced KMC sampling}

Using a KMC algorithm to determine equilibrium properties can be very inefficient if states of very low probability need to be crossed frequently to obtain good statistics. Here, we address this issue by devising a simple biasing scheme, similar in spirit to umbrella sampling, that allows us to relate equilibrium properties computed using biased transition rates with equilibrium properties of the reference system. Supposing that the state space of the system is divided in windows according to a certain order parameter $\boldsymbol{Q}$, if transition rates are modified according to
\begin{equation}
k(\vec{y}|\vec{x})\rightarrow \frac{W[\boldsymbol{Q}(\vec{y})]}{W[\boldsymbol{Q}(\vec{x})]}k(\vec{y}|\vec{x}),
\label{eq:4}
\end{equation}
then, assuming detailed balance in the underlying rates $k(\vec{y}|\vec{x})$, the probability distribution $P'(\boldsymbol{Q})$ of the modified system relates to those of the reference system $P(\boldsymbol{Q})$ through
\begin{equation}
\frac{P(\boldsymbol{Q}_a)}{P(\boldsymbol{Q}_b)}=\frac{P'(\boldsymbol{Q}_a)}{P'(\boldsymbol{Q}_b)}\left[ \frac{W(\boldsymbol{Q}_b)}{W(\boldsymbol{Q}_a)}\right]^{2},
\label{eq:enhanced_kmc}
\end{equation}
where $\boldsymbol{Q}_a$ and $\boldsymbol{Q}_b$ denote any two points of the order parameter space. In practice, the weights $W(\boldsymbol{Q})$ were first adjusted so that every value of the order parameter could be observed in the simulations, regardless of the initial state. From the first estimate of the probability distribution $P(\boldsymbol{Q})$ a new set of weights were derived from Eq.~\ref{eq:enhanced_kmc} by fixing $P'(\boldsymbol{Q})=\textnormal{const}$. The simulations were performed again  using the new set of weights and the method repeated until a final set of weights $W(\boldsymbol{Q})$ is found, such that any two values of the distribution $P'(\boldsymbol{Q})$  differ by no more than 10\%.

\subsection{\label{sec:level2} Neglected effects}

In our coarse-graining scheme we ignore several effects that we discuss below.
\subsubsection{Long-range bridging effects} 

Firstly, {\em long-rage bridging effects}  can be relevant if a tile either incorporates or dissociates from a position in the assembly that acts as a bridge connecting two distant sections (an example is presented in Section S.VII). To gauge how important these cases are, we performed independent simulations using two different versions of the model. In the first version, the rates of addition or removal of ``bridging-tiles" were assumed to be the same as for an otherwise equivalent environment in which both sections of the assembly are constrained, and the local approximation can be applied. As an example, the rates of tile association or tile removal for the local environment  in Section S.VII were assumed to be the same as the rates for the environment in Fig.~\ref{fig:le_examples}b. In the second version of the model, the addition (and removal) of ``bridging-tiles" was prohibited altogether. All the simulations performed were repeated using both versions of the model. As the entropic penalty for bringing together both sections of the structure is not accounted for in the first version of the model, bridge formation is artificially favored compared to a (hypothetical) more complete version of the model where non-local effects are treated correctly. Nevertheless, in the simulations performed with the first version of the model bridge formation occurred so sporadically that the overall time scales and thermodynamic properties barely varied with relation to those observed using the second version of the model. Thus, it was concluded that bridge formation is rare enough that it can be ignored. The results presented throughout this work were obtained using the second version of the model, where bridge formation was prohibited. 

We note that throughout this work we study assembly under moderate supercooling conditions, where we find growth to occur through a series of compact intermediates with a high number of bonds between components. However, at higher supercooling the shape of intermediates is more ramified as tile association tends to be more irreversible. In this regime,  bridge formation might become more relevant to the assembly pathways, and thus require an explicit treatment.

\subsubsection{Multiple copies of the same tile species in a structure}
 Secondly, we  \textit{ignore states in which multiple copies of the same tile species are present in a structure}, each copy bonded through a different set of (programmed) binding domains. During tile association, as the first contacts begin to form, we expect further bonding to occur relatively quickly (it is a localised first-order process) compared to the association of a second tile, which is a second-order process whose rate we expect to be much slower at strand concentrations typical of experiment. Of course, at sufficiently high strand concentrations this approximation is likely to begin to break down, but we expect this to only occur well above the range relevant to experiments.

\subsubsection{Local environment approximation} 

Thirdly, we tested the {\em local environment approximation} for a few simple cases with 1 or 2 binding domains where we included a much larger number of strands in the assembly. The differences in the respective free-energy profiles were very small, and so we expect this to be a good approximation. 

\subsubsection{Misbonding} 
 Fourthly, we ignore {\em misbonds} in our kinetic model, assuming interactions respect the domain-level description, and only occur between domains that are intended to be complementary.    This approximation greatly simplifies our calculations.   In Section S.III we sketch an upper bound on the average effect of misbonds.  We briefly summarise the calculation below.  

Firstly, we attempt to identify an upper bound for the typical binding probability of an unintended strand to an available ``site" within an assembly. We assume that this site has 42 available bases (in general, an overestimate) and further treat these bases as if they were presented by a single contiguous strand, rather than on separate tiles (again, overestimating the stability of bonding by neglecting entropic penalties due to the complexity of the environment). We generate a representative set of 42-base sequences for the site by using the sequences of the 334 tiles themselves.

For each of these representative sites, we consider the total  binding free energy of strands that are not intended to bind to any of the domains in the site.  We perform this calculation using the NN model,\cite{SantaLucia1998} considering the contribution to the ensemble of all possible base-pairing configurations with either one or two base-paired sections of 2 base pairs or more. 

At the relevant temperatures, we find that the binding free energies for a given strand and a given site are significantly higher than if the strand is designed to bind with that site, with a few extreme sequences with free-energies of association comparable to an $N_b=1$ local binding domain (see Fig. S2). Averaging over sequences, we can compare the relative probabilities that an available site will bind to a correct strand or an incorrect one in equilibrium. We find that at the range of temperatures where self-assembly takes place, our upper bound on the misbond probability falls below that of an $N_b=1$ local binding domain (see Fig. S3), which is the weakest of the correct binding motifs. Because the binding probability at equilibrium of tiles with a higher connectivity $N_b>1$ is much higher than for tiles with a single binding domain (Fig. S3), we conclude that misbonding has a negligible effect on assembly at equilibrium.
 
We also estimate the effect of misbonding on dynamics by employing a simplified variant of our assembly model where correctly bonded tiles are not allowed to dissociate from the structure once bonded, but incorrectly bonded tiles can associate and dissociate with rates that were derived using our upper bound on the average probability for misbonding. At higher temperatures ($45^\circ\text{C}-50^\circ\text{C}$),  we find assembly times  during an isothermal assembly protocol to be largely insensitive to incorrect bonding (see Fig. S4). At lower temperatures (typically $T<40^\circ\text{C}$) we find the assembly dynamics for an isothermal protocol to be increasingly slowed down due to misbonding. 

However, it should be kept in mind that at these low temperatures, our approach of studying growth of a single cluster is likely to be inappropriate, and the rapid nucleation of many clusters that are unable to grow to completion because of monomer starvation is likely to have a more dominant negative effect on the assembly dynamics than misbonding.  

Interestingly, the upper bound on misbonding does not appreciably change if we simply take a set of random strands (see Fig. S2).  It should be kept in mind that typical DNA design spaces are  enormous\cite{louis2016contingency} --  for this system there are $1.9 \times 10^{25}$ different possible sequences of length $42$ -- and so it is not hard to see that the probability of finding two random sequences that have a significant amount of complementarity is small. For example, the probability of two random sequences of length $42$ having a complementary section 10-bases long, the typical length of the bonding domains in the assembly, is $6.2\times 10^{-4}$ (see Section S.V). Most complementary motifs are short, and it is rare to find one more than about 5 bases long between any two strands in a set of 334 randomly generated tiles.  In the experiments~\cite{PengYin2d} it was also found that randomly picking sequences (that is not trying to design out misbonding) did not make the SST system more likely to aggregate, which accords with our findings above.   For the SST assembly system, it appears to be the case that the  exponentially large design space  means that misbonding,  which in the past was often thought to be a very significant barrier to the success of an addressable assembly method, can be avoided quite straightforwardly.

\begin{figure}[h]
    \centering
    \includegraphics[width=0.45\textwidth]{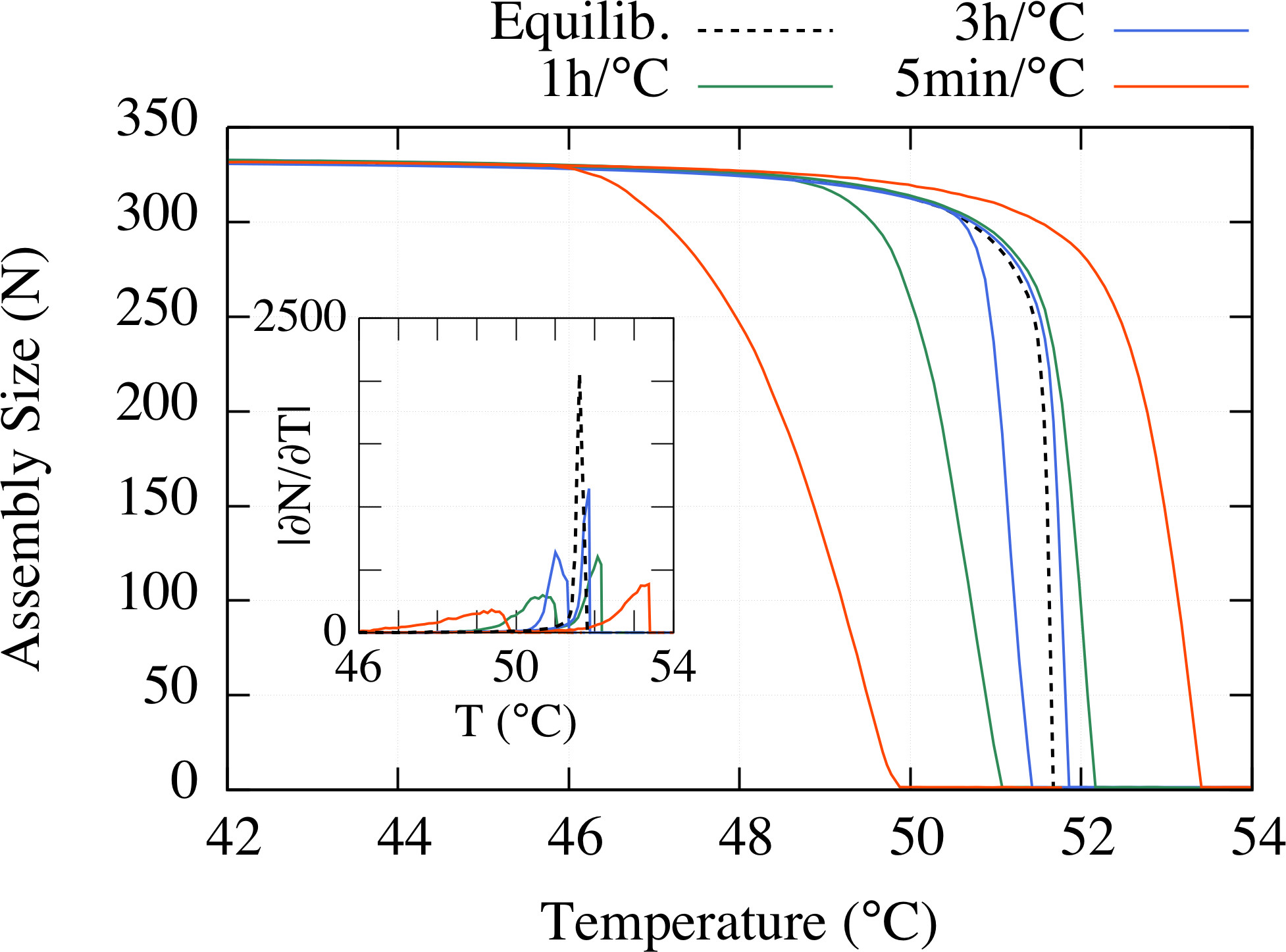}
    \caption{Annealing and melting curves using different cooling/heating rates. Also shown in the figure (dashed black curve) is the assembly size at equilibrium as a function of temperature. Inset: derivatives with respect to temperature of the curves in the main plot. The assembly peaks (inset) have maxima at $51.0^{\circ}\text{C}$ for $3\text{h}/^\circ\text{C}$, $50.7^{\circ}\text{C}$ for $1\text{h}/^\circ\text{C}$, $49.4^{\circ}\text{C}$ for $5\text{min}/^\circ\text{C}$ and $51.64^{\circ}\text{C}$ for equilibrium. The melting peaks (inset) have maxima at $51.9^{\circ}\text{C}$ for $3\text{h}/^\circ\text{C}$, $52.1^{\circ}\text{C}$ for $1\text{h}/^\circ\text{C}$ and $53.4^{\circ}\text{C}$ for $5\text{min}/^\circ\text{C}$. }\label{fig:assembly_rate}
\end{figure}

\begin{figure*}[ht]
    \centering
    \includegraphics[width=1.0\textwidth]{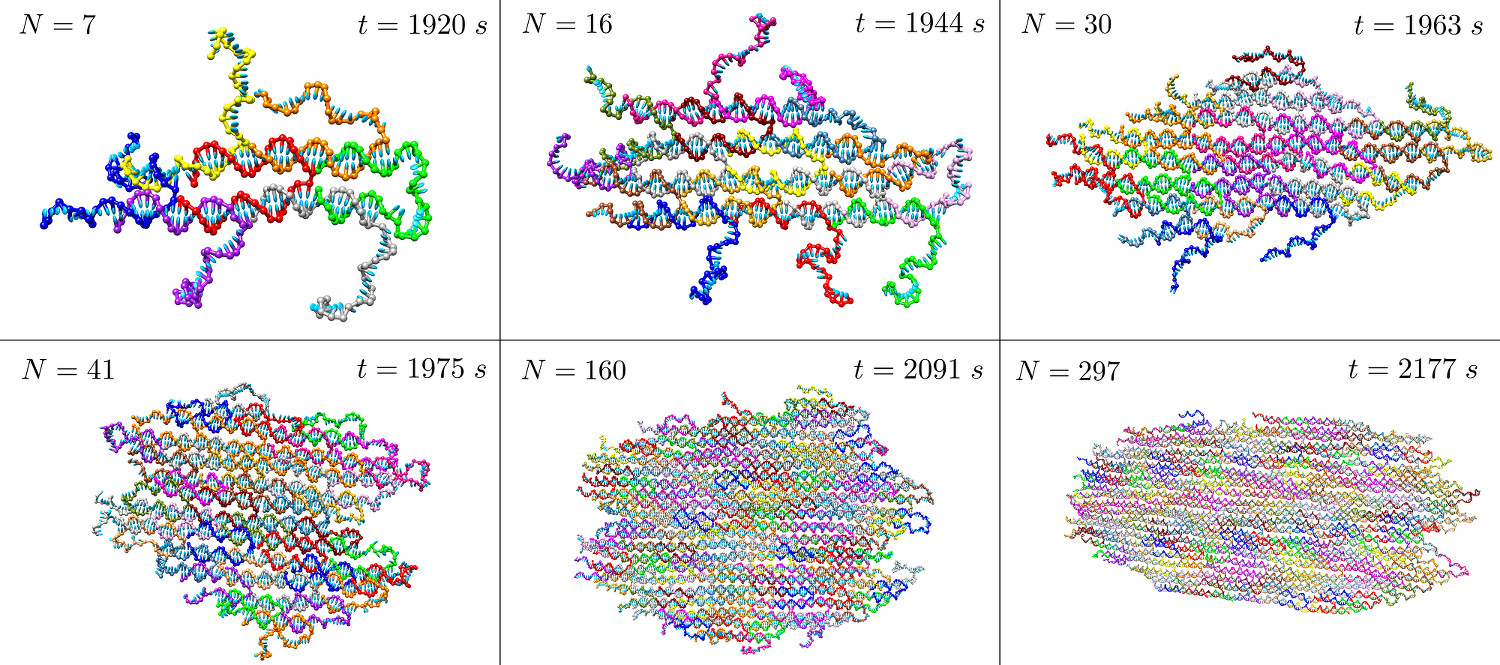}
    \caption{Illustrative oxDNA configurations for the states of the system during an isothermal trajectory at $T=51^{\circ}C$. Note that for this isothermal protocol complete assembly is never reached, but rather equilibrium is for an average size of 297 tiles. Lower temperatures are needed for full assembly.}
\label{fig:snapshots}
\end{figure*}

\section{\label{sec:level1}Results}
 
 \subsection{Assembly with a thermal ramp protocol} 

The oxDNA model predicts a critical temperature of $T_\text{c}=51.64^\circ\text{C}$, defined as the temperature where it first becomes thermodynamically favourable for a large assembly to form (the details of how $T_c$ is calculated are given in Section III.C). 

In the original experiments to which we are comparing ~\cite{PengYin2d,SobczakDNAfolding} a thermal annealing protocol with steps of $1^\circ\text{C}$ was used.  For a cooling protocol of $3 \text{h}/^\circ\text{C}$, the maximum assembly rate was measured at $53^\circ\text{C}$ by Sobzak {\it et al.}.\cite{SobczakDNAfolding} Given the resolution of the $1^\circ\text{C}$  steps, we estimate that for experiments, the critical temperature is somewhat above $53^\circ\text{C}$, possibly close to $54^\circ\text{C}$, where structure growth was first observed in the assembly mixture.  A full comparison with the dynamics of assembly in the experiments would need to reflect the correct ensemble, which means a fixed number of tiles, so that the change in concentration of tiles with assembly growth can be taken into account.    Nevertheless, the critical temperature observed in single target simulations corresponds to the onset temperature where it first becomes thermodynamically favourable for structures to form in an assembly scenario, and so should be accurately represented by our single-target system.   The agreement with our calculated critical temperature is remarkable given that no adjustable parameters are used in the model that affect its thermodynamic predictions (note that the forward reaction constant $k_+^0$ only affects the time scales). In Kelvin, the appropriate temperature scale for statistical mechanical models, this agreement is better than 1$\%$. Note that this agreement also implies that the oxDNA model is accurately capturing the magnitude of the barriers to multiple domain formation.

Keeping in mind that our current single-target assembly approach does not take into account the effects of a fixed number of monomers as in experiments, it is nevertheless instructive to study the effect of cooling and heating protocols.    Starting at $T=60^\circ\text{C}$, we simulate an annealing process by decreasing the temperature in discrete steps (assumed to be instantaneous) of $\delta T=0.1^\circ\text{C}$ until the temperature reaches $T=40^\circ\text{C}$. This is then followed by a melting protocol where the temperature is increased back to $T=60^\circ\text{C}$ using the same temperature step. For each cooling/heating rate used, we produced $10^{4}$ independent trajectories, each initiated from the assembly state $x_{\mu}=\delta_{\mu \nu}$, where $\nu$ was selected randomly from the set of 334 different tile species $(\nu \in [1,334])$.  
In Fig.~\ref{fig:assembly_rate} we present the average assembly sizes observed at the end of each temperature step for both annealing and melting. Also shown in Fig.~\ref{fig:assembly_rate} are the (finite) derivatives of both annealing and melting curves (inset), and the equilibrium assembly size. For each of the  cooling/heating rates used ($15\text{min}/^\circ\text{C}-3\text{h}/^\circ\text{C}$), we find differences between the peaks for assembly and melting, i.e. the system is not at equilibrium and exhibits hysteresis. Interestingly, melting is found to occur closer to $T_c$ than assembly. Also, we find the melting peaks to be sharper than the assembly peaks (Fig.~\ref{fig:assembly_rate} - inset), which suggests that melting has a higher degree of cooperativity than assembly.

\begin{figure}[h]
    \centering
    \includegraphics[width=0.45\textwidth]{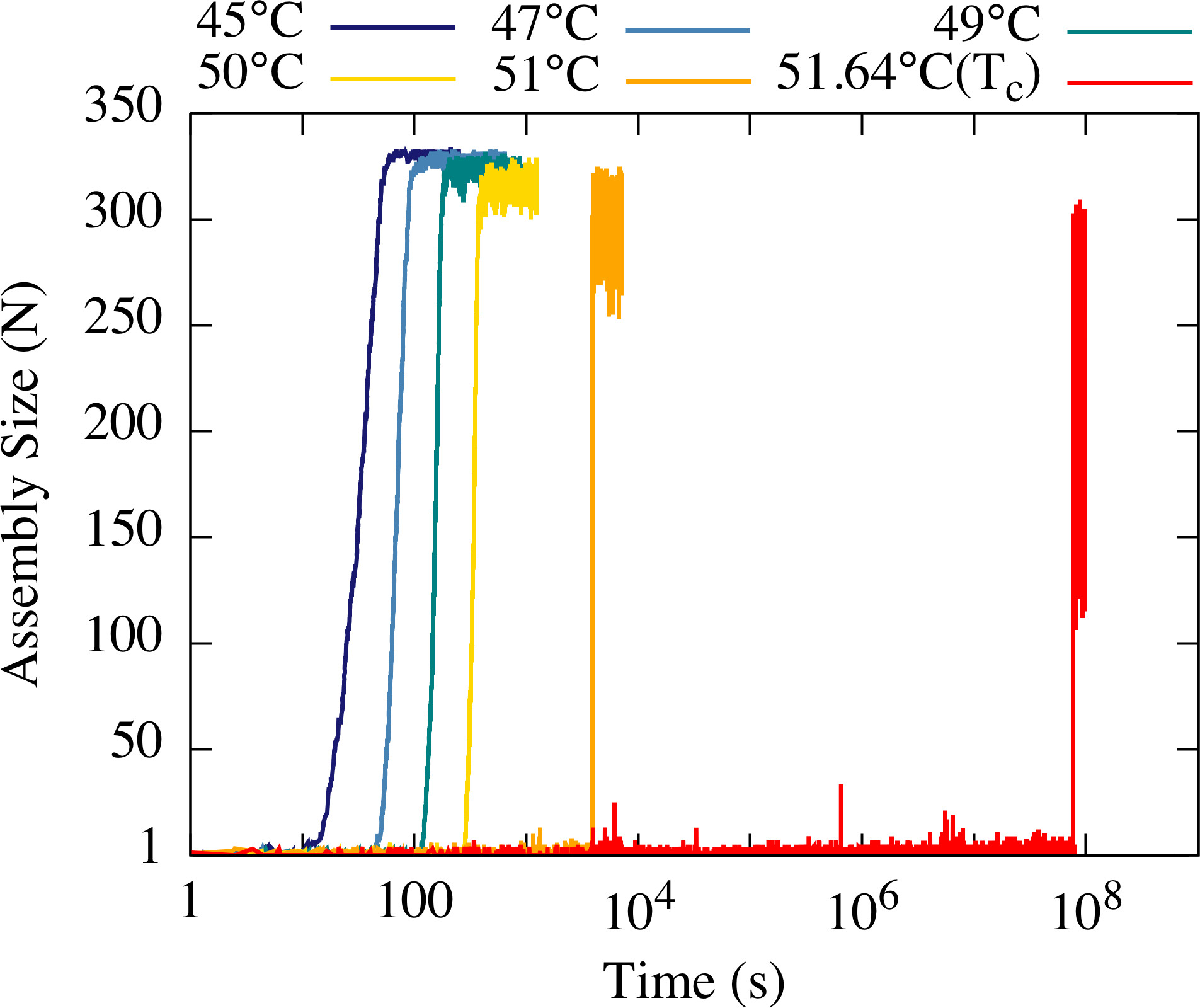}
    \caption{Assembly size versus time curves from isothermal simulations at different temperatures. }\label{fig:trajectories}
\end{figure}

 \begin{figure}[t]
    \centering
    \includegraphics[width=0.45\textwidth]{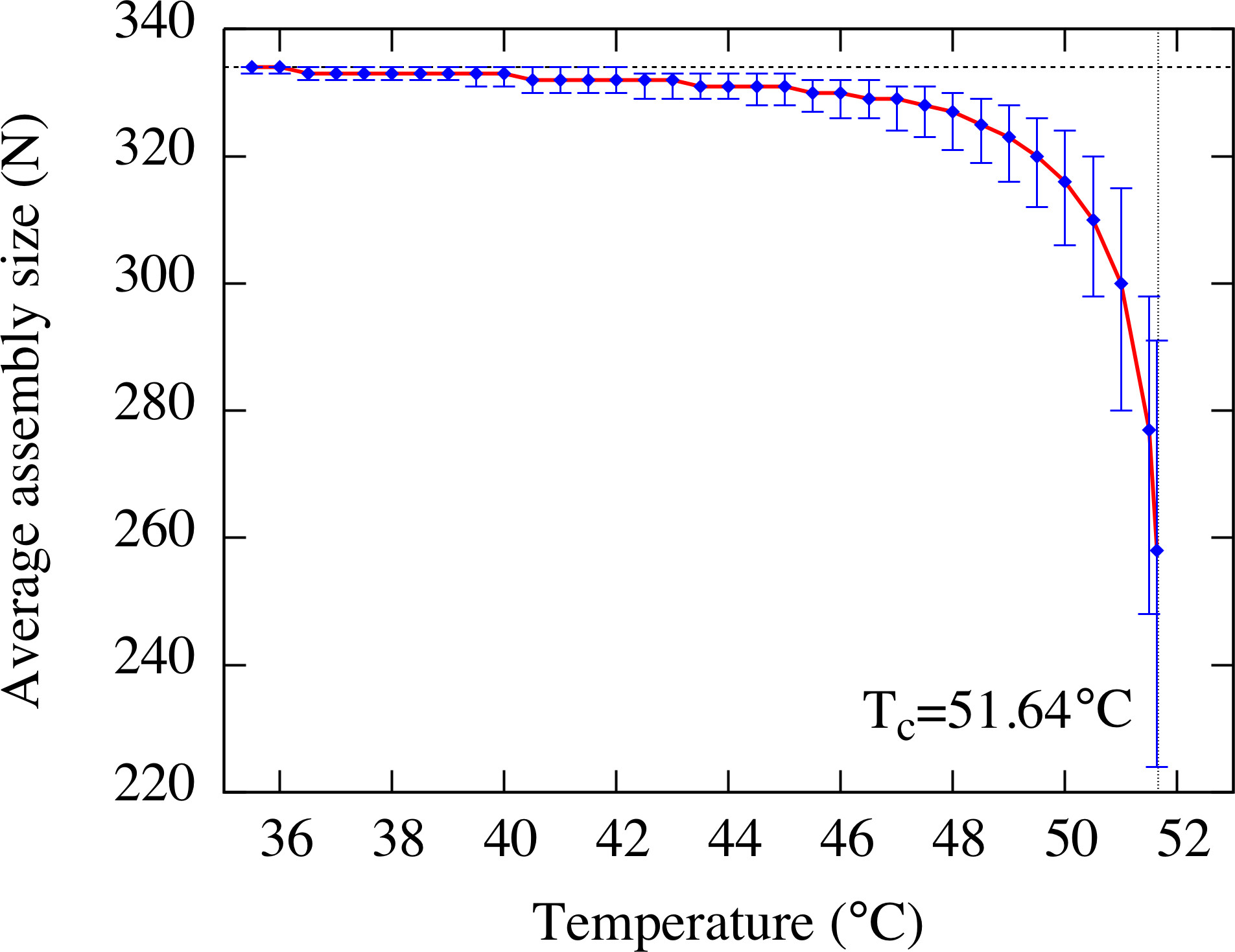}
    \caption{Equilibrium assembly size (blue diamonds) as a function of temperature, and the respective size fluctuations. The vertical bars indicate the range of sizes that are observed at equilibrium with $95 \%$ probability.  Note that the fluctuations in size are not symmetric, fluctuations  are larger to smaller than average sizes.}\label{fig:assembly_sizes}
\end{figure}

\subsection{Isothermal assembly} 
We also simulated  isothermal assembly by producing $10^{3}$ uncorrelated trajectories at several (constant) temperature values in the range $45^{\circ}\textnormal{C}-51^{\circ}\textnormal{C}$. In Fig.~\ref{fig:snapshots} we show typical oxDNA representations of states of the assembly taken at different time slices from one of the trajectories at $51^{\circ}\textnormal{C}$. The time evolution of assembly yield, presented in Fig.~\ref{fig:trajectories}, shows a number of interesting features. Firstly,  at lower temperatures the size of assembly increases quickly after the trajectories are started. However, at  higher  temperatures, the system exhibits increasingly longer waiting times in which the assembly does not grow beyond a few tiles, followed by a relatively quick growth towards large stable intermediates. Note also the large range of waiting times in Fig.~\ref{fig:trajectories}.  For example, at the critical temperature, we measure a waiting time on the order of several years. A small drop in temperature to $51^{\circ}\textnormal{C}$ yields much faster assembly on the order of hours. By $49^{\circ}\textnormal{C}$ the waiting time is reduced  to just a few minutes.  All the above is consistent with a nucleation barrier to assembly that rapidly decreases in size with decreasing temperature. 

  At the critical temperature of $51.64^{\circ}\textnormal{C}$, the final structure has an average size of around 260 tiles, with large fluctuations (See Figs.~\ref{fig:trajectories}~and~\ref{fig:assembly_sizes}).  As the temperature is dropped,  the assembly grows to a larger  final state, with smaller fluctuations. Fully assembled structures involving all 334 strands only appear with a probability greater than 50$\%$ at temperatures of  $36^\circ$C and below.

It should be kept in mind that these calculations were performed for single target assembly in a constant background concentration of unbound tiles, and thus, structure growth is not hindered by the consumption of monomers by competing nuclei. However, we can infer from our isothermal curves (Fig.~\ref{fig:trajectories}) that there is a narrow temperature window, between $\sim 49^{\circ}\textnormal{C}$ and $51.64^{\circ}\textnormal{C}$, where monomer starvation will be avoided in an isothermal assembly scenario, as there is a clear timescale separation between waiting times and growth.

\subsection{Free energy of assembly and nucleation barriers}

\begin{figure}[h]
    \centering
    \includegraphics[width=0.45\textwidth]{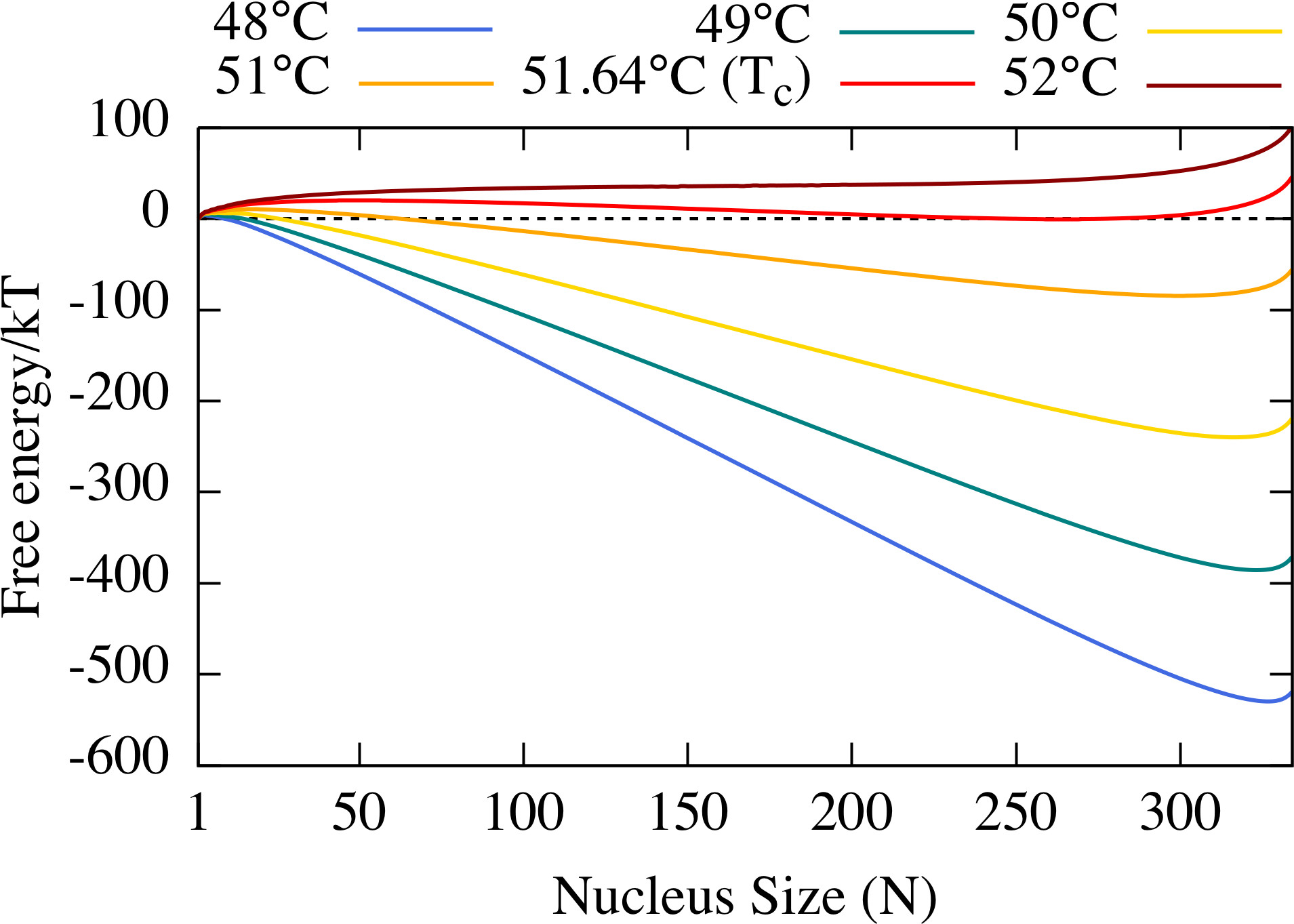}
    \caption{Free-energy profiles as function of assembly size for different temperatures. }\label{fig:free_energy_large}
\end{figure}

\begin{figure}[h]
    \centering
    \includegraphics[width=0.45\textwidth]{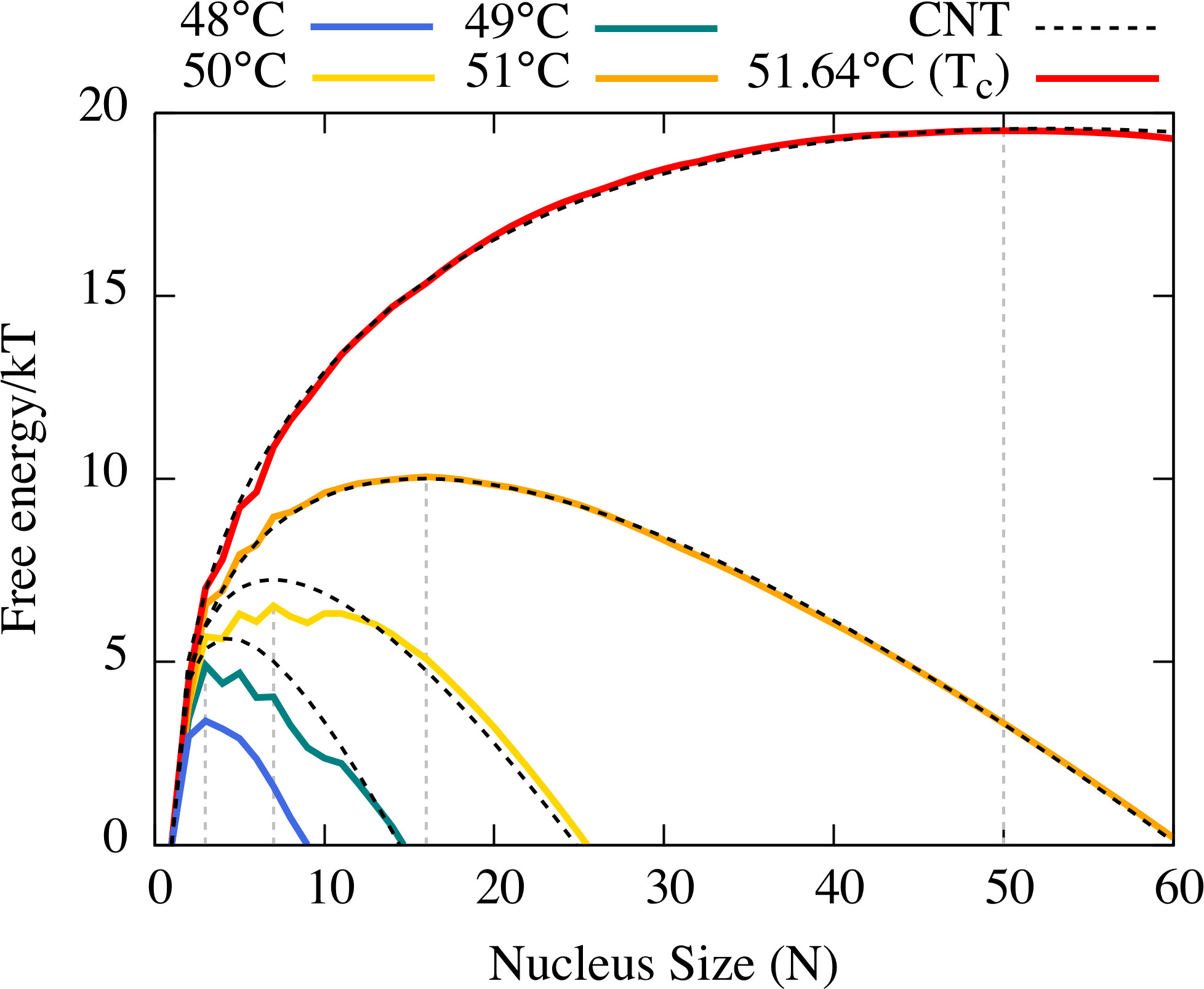}
    \caption{  Higher resolution plot that shows the nucleation barriers for different temperatures, and the respective CNT fits (obtained using the free energy values for $N<200$). The dashed vertical lines indicate the critical nucleus sizes for the nucleation barriers simulated: $N^*=3,7,16,50$ for $T=49^\circ\text{C},50^\circ\text{C},51^\circ\text{C},51.64^\circ\text{C}$, respectively. }\label{fig:CNT_fit}
\end{figure}

In order to further understand the behaviour observed in Figs.~\ref{fig:trajectories} and~\ref{fig:assembly_sizes},  free-energy profiles were obtained using the  biased KMC simulation  method described in Section II.C. In the calculations, we used the size of the assembly $(N)$ as the biasing parameter (see Section II.C). The results (Figs.~\ref{fig:free_energy_large} and~\ref{fig:CNT_fit}) confirm the existence of a nucleation barrier that increases quickly as the temperature approaches the critical temperature from below.  The critical nucleus size for each temperature was confirmed by using unbiased simulations to verify that for trajectories initiated at the critical nucleus $50\%$ go on towards the formation of a large assembly, whilst in the other $50\%$ the nucleus shrinks back down.

The nucleation behaviour changes rapidly over just a few degrees.  The largest critical nucleus is $N^*=50$ at the critical temperature of $T=51.64^\circ\text{C}$, but by $T=51^\circ\text{C}$ it has already dropped to $N^*=16$. Just a few degrees lower, at $49^\circ\text{C}$, the critical nucleus is only $N^*=3$, so that the nucleation time will be very fast and no longer significantly limit the rate of assembly.

The free energy exhibits a minimum at an assembly size that is less than that of the full 334-strand structure, and the size at this minimum grows with decreasing temperature. This behaviour arises due to the finite size of the target structure. Close to completion the number of possible combinations of tiles for a given assembly size decreases rapidly, reaching $1$ for the complete assembly. This entropic effect is characteristic of finite-size assemblies and is absent, for example, in conventional crystal nucleation, where growth is solely limited by supersaturation.

This feature of the free-energy curves explains the final assembly sizes observed in Figs.~\ref{fig:trajectories}~and~\ref{fig:assembly_sizes}.  The changing location of this minimum  also explains  why a thermal cooling ramp is needed for full assembly of the SST system.\cite{lattice_model_wang2}  At temperatures where the nucleation barrier is large enough that a separation of timescales emerges between nucleation (slow) and further growth (fast), the final equilibrium structure is smaller than the fully formed structure.  At lower temperatures, where a fully formed structure is thermodynamically stable, the nucleation barrier is so small that  nucleation will be fast compared with growth, leading to many partially formed structures and to monomer starvation if assembly is conducted under isothermal conditions. Thus, an annealing process is needed that enables controlled nucleation and growth towards near completion at high temperature, followed by the growth towards completion as the temperature is decreased.

\begin{figure}[h]
    \centering
    \includegraphics[width=0.45\textwidth]{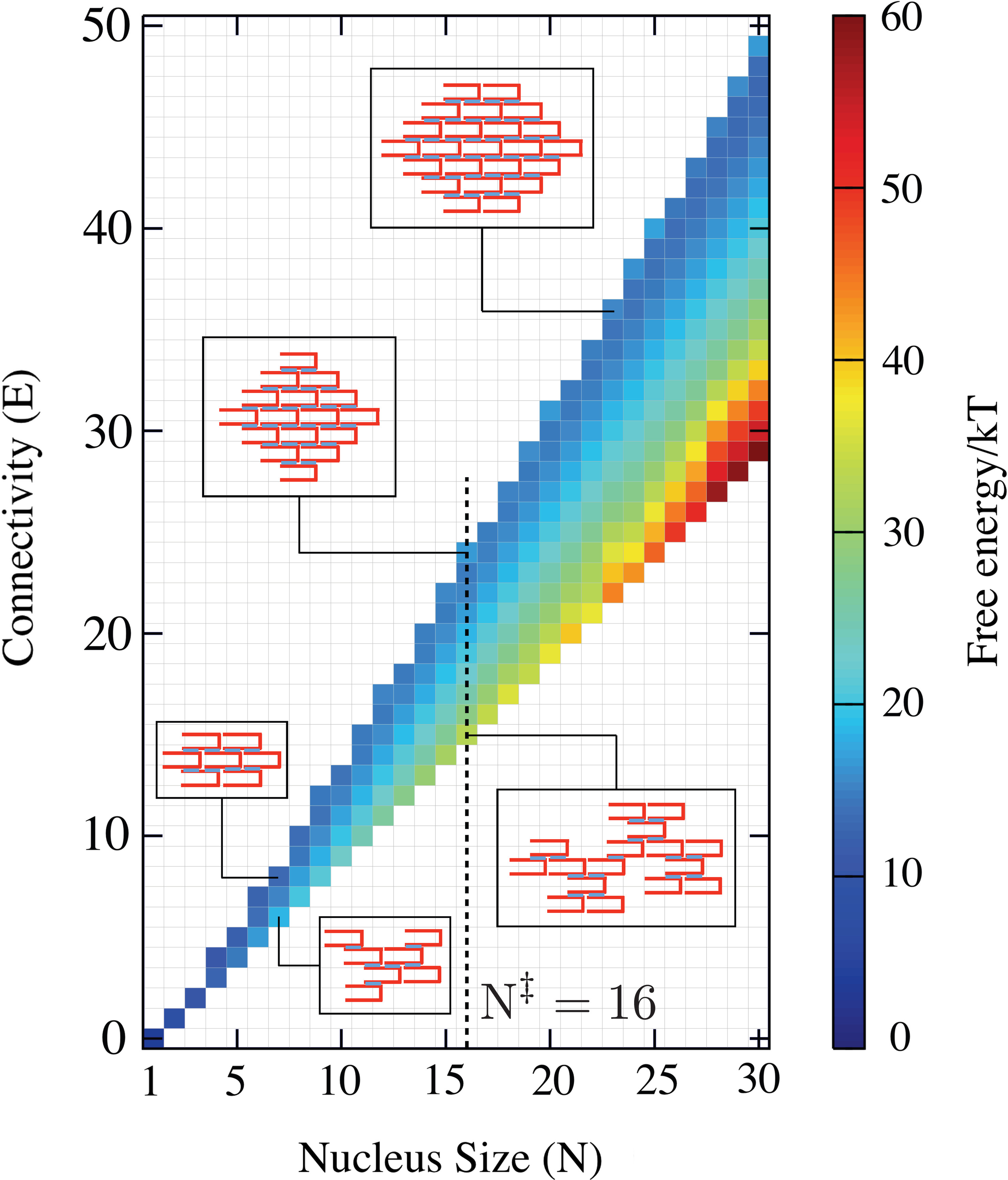}
    \caption{Free-energy landscape for assembly, as a function of the size ($N$) and number of bonded domains ($E$) between tiles, at $T = 51^\circ\text{C}$. The bubbles show typical examples of assembly states  representative of the respective macrostate ($N,E$). The critical nucleus size at $T = 51^\circ\text{C}$ is $N=16$ (dashed vertical line).}\label{fig:free_energy_N_E}
\end{figure}

\begin{figure}[h]
    \centering
    \includegraphics[width=0.45\textwidth]{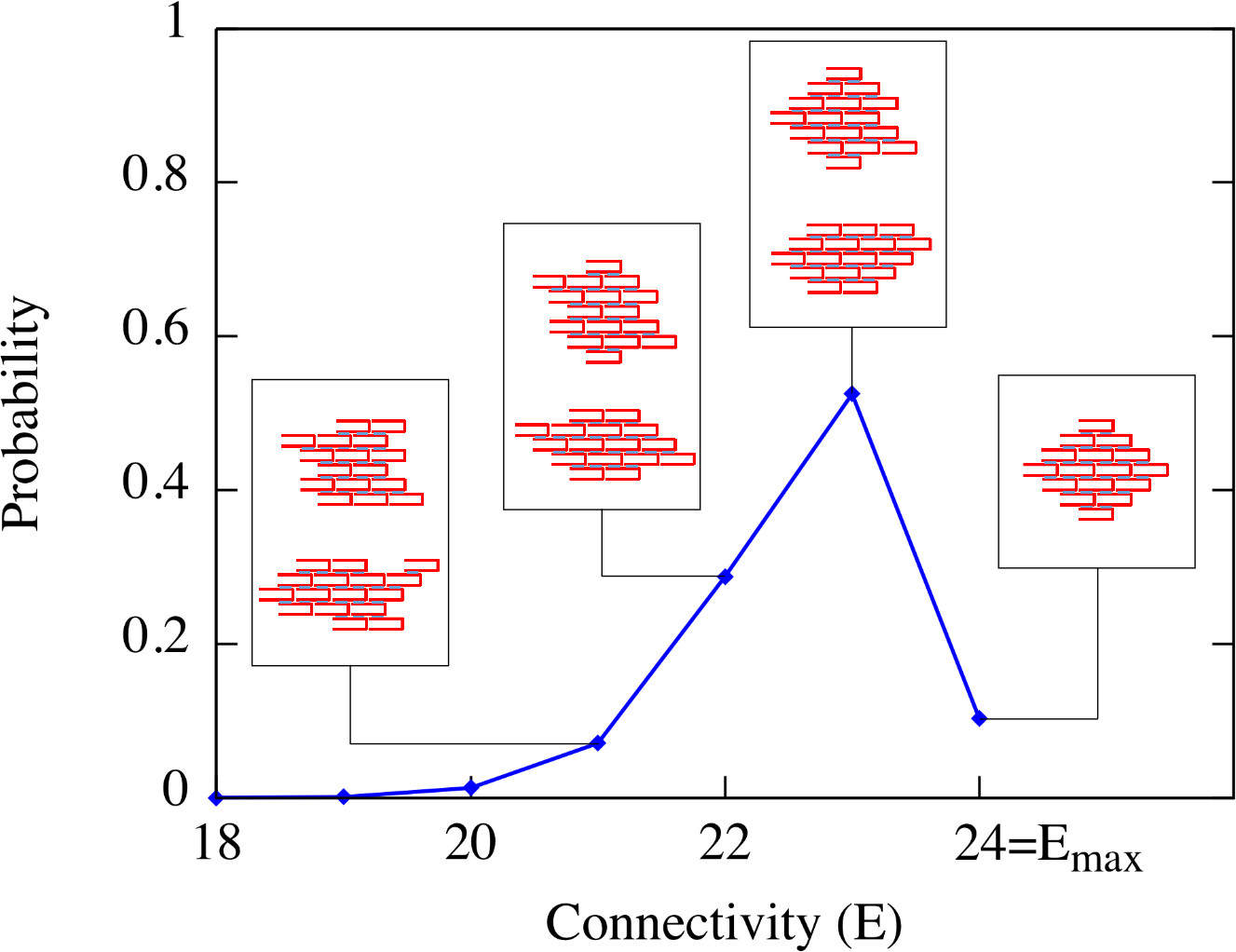}
    \caption{Probability of a nucleus as a function of the number of bonded domains ($E$) between tiles for structures with $N=16$ components (critical size at $T = 51^\circ\text{C}$). Above the curve are depicted examples of assembly states, representative of states  $(N=16,E)$ with different connectivities.}\label{fig:P_E_51}
\end{figure}

To characterize the assembly reaction pathways in more detail, biased simulations were performed at several temperatures to determine the free-energy landscape as a function of the size $(N)$ and connectivity $(E)$ of an assembly, the latter defined as the number of bonded domains in the structure. In the calculations, we used the size of the assembly $(N)$ and the connectivity $(E)$ as the biasing parameters (see Section II.C).
As can be seen in Fig.~\ref{fig:free_energy_N_E}, for a fixed size, higher connectivity is generally favoured, as this means a more favourable enthalpy.  But entropy also plays a role and disfavours structures with full connectivity, so the most likely connectivity typically lies just below the maximum.  In Fig.~\ref{fig:P_E_51}, we illustrate this entropic effect for the $N=16$ critical nucleus size at $T=51^\circ\text{C}$, and for the $N^*=50$ critical nucleus size at the critical temperature ($T_c=51.64^\circ\text{C}$) in Fig. S.IV. This entropic effect is increasingly relevant for larger nuclei, because the number of possible alternative structures with connectivity just below that of the maximum increases with size. 

\begin{figure}[h]
    \centering
    \includegraphics[width=0.45\textwidth]{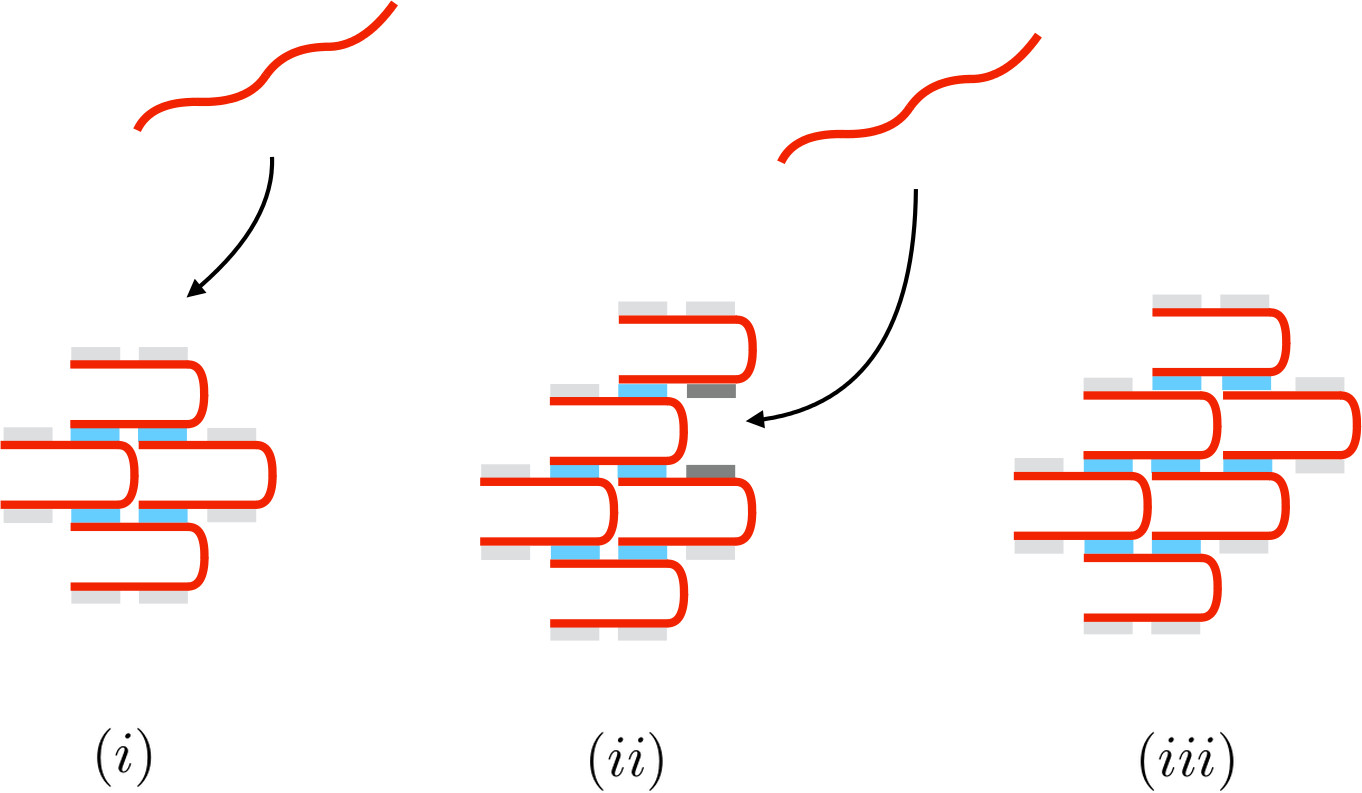}
    \caption{Cooperative growth of a closed motif. Closed motifs are structures where each tile is connected with the remaining through at least two binding domains, and the binding sites at the interface have a single domain available for binding. In order for a closed motif to grow, one tile first needs to bind through a single binding domain $(i)$, that creates one binding site with two domains available for binding ($ii$-dark grey), that have a chance of being occupied if a second tile further associates, with extra stabilization effect $(iii)$.}\label{fig:cooperative}
\end{figure}

For example, if we look closely at the free-energy profiles in Fig.~\ref{fig:CNT_fit}, we see some small features at small nucleus size with nuclei with $4$, $6$ and $9$ strands showing deviations to lower free energy, but with the curves quickly becoming smooth beyond this point. These sizes correspond to complete $2\times 2$, $2\times 3$ and $3\times 3$ rhomboids. For example, as illustrated in Fig.~\ref{fig:cooperative}, the extra stability of the $N=6$ cluster is because the last strand that binds to form this cluster gains two binding domains in contrast to the previous or next strand to add to the growing assembly, which can only bind with one domain. 

In summary, these results show that, for all but the smallest nuclei,  the most likely assembly pathways are through a set of highly connected structures with a near rhomboidal shape, but which are not the the most connected structures.

This behaviour contrasts with previous theoretical work,\cite{lattice_model_wang} using a more coarse-grained approach where it was found that  maximally connected structures have the lowest free energy and are the most likely to be formed, and where the free-energy profiles show much stronger features associated with sizes where cycles of connections can be completed, that persist to significantly larger sizes.  The reason for these differences with the current work is the comparative size of the free-energy barriers associated with initiating the second and subsequent domains. In the previous patchy particle models, these barriers are significantly lower, leading to a greater stabilization for those strands that bind with two domains and hence to a stronger preference both for nuclei that are fully-connected, and for sizes for which compact, ``closed'' motifs are possible. By contrast, when the full entropic cost of initiating the second and subsequent domains is properly accounted for, as we do here, the free-energetic advantage of strands that can bind with multiple domains is significantly diminished.

We next consider whether the main features of the nucleation barrier can be captured by classical nucleation theory (CNT), which models the competition between the addition of monomers, which lowers the free energy and the  formation of a surface (or line in 2D), which costs free energy.  For a circular structure free of anisotropies,  this leads to a simple expression 

\begin{equation}
 \Delta G(N)=- |\Delta \mu |N+ \Lambda N^{1/2},
 \label{eq:CNT_fit}
\end{equation}
where $N$ is the number of monomers forming the structure, $\Delta \mu <0$ is the chemical potential difference between monomers forming the structure and monomers in the reservoir, $\Lambda=2\pi^{1/2}\rho_{s}^{-1/2}\lambda $ where $\lambda $ is the line tension, and $\rho_{s}$ is the  number density of monomers forming the structure. 

To estimate the number density of monomers we simulate a fully assembled structure (Fig.~\ref{fig:membrane}) using molecular dynamics simulations with oxDNA2,\cite{snodin2015introducing} an extension of the original oxDNA model that allows for a better description of the structural properties of large (kilobase-pair) DNA nanostructures. The simulations were performed at  $T = 25^\circ\text{C}$ and after a period of equilibration, we measured the average distance between the centres of mass of contiguous tiles. We then estimate the area occupied by each tile by assuming that contiguous tiles can be considered to be contained in the same plane. Using this method, we obtain the value $\rho_{s}=5.82\times 10^{16}\, \textnormal{m}^{-2}$ for the surface density of monomers.

We use the CNT expression to fit our free-energy curves for assemblies with  up to 200 monomers; we exclude larger sizes as CNT is not able to capture the entropic effects that lead to a minimum in the free energy near to the fully-assembled state. Note also that in order for the CNT fit to coincide with our free-energy curves at $N=1$ (where we defined $G(1)=0$ for all temperatures) we use the fitting variable $N'=N+1$ in Eq.~ \ref{eq:CNT_fit}.

In Fig.~\ref{fig:CNT_fit}, we show that the nucleation barrier for  several temperatures is fit well by the classical CNT expression with the values  of the parameters $|\Delta \mu |$ and $\lambda$ given in Table~\ref{table:1}.  Of course this simple expression will not capture the size-specific features observed for very small nuclei, and the minimum at larger sizes, but it is nevertheless remarkable how well it fits the overall curve.   As is found for many other systems,  the surface (line) tension does not change much with temperature. Instead, the  rapid drop in the barrier height with temperature is mainly caused by the rapid change in $\beta \Delta \mu$.

\begin{table}[ht]
\centering 
\caption{CNT fit parameters} 
\begin{ruledtabular}
\begin{tabular}{c c c} 
$T (^\circ\text{C})$ & $\beta |\Delta \mu |$ & $\lambda (\textnormal{pN}) $ \\ [0.5ex] 
\hline 
$49$ & 1.26 & 1.42  \\ 
$50$ & 0.99 & 1.61 \\
$51$ & 0.67 & 1.51 \\
$51.64\, (T_c)$ & 0.36 & 1.58 \\ [1ex] 
\end{tabular}
\end{ruledtabular}

\label{table:1} 
\end{table}

Although experimental measurements on the line tension of DNA assemblies have not to our knowledge been made, it is interesting that  predictions are fairly similar to typical experimental measurements on various lipid bilayers  which range from ($\lambda =1\, \textnormal{pN}$\cite{line_tension_exp1} to $\lambda =4\, \textnormal{pN}$\cite{line_tension_exp2}), as well as predictions using Monte-Carlo simulations for a coarse grained lipid model ($\lambda =3\pm 2\, \textnormal{pN}$\cite{line_tension_mod1}).   

\section{\label{sec:level1}Conclusions}

In this work, we develop a two-step coarse-graining approach that uses extensive thermodynamic calculations with a nucleotide-level model of DNA, to parametrise a coarser kinetic model that is able to reach the time and length scales needed to describe explicitly the self-assembly of a two-dimensional $334$-strand SST structure first produced experimentally by the Yin group.\cite{PengYin2d} The resulting kinetic model provides a detailed description of the assembly trajectories and, without using adjustable parameters, it predicts a transition temperature which is consistent with previous experimental observations. The model used to calculate the thermodynamics of tile association, oxDNA, describes explicitly the polymeric degrees of freedom of DNA, which makes it more suitable to capture the entropic penalties associated with initiation of a second and subsequent domains during tile association than previous patchy-particles models.\cite{lattice_model1,lattice_model2} 

 For the SST system,  we show that there exists a narrow temperature regime where the nucleation barrier is low enough to be crossable on experimental time-scales, but large enough to allow a time-scale separation between growth and nucleation. At the same time,  the exponentially large design space for DNA means that misbonds do not overwhelm correct assembly at experimentally relevant temperatures.  

We find that the critical nuclei are made up of  an ensemble of structures that are highly, but typically not maximally, connected.  
The nucleation barrier near the temperature where assembly first occurs is surprisingly well described by CNT, with the fits providing reasonable estimates for the surface (line) tension and chemical potential difference.  At lower temperatures and for smaller nuclei, the full model exhibits structure  due to cooperative effects not included in CNT. In contrast to systems with only a few components, the only way to achieve full assembly is to use a cooling protocol, where the system first crosses the nucleation barrier to a partially formed assembly, and further cooling stabilises the full assembly.   
While our calculations are quantitative for a specific 334 strand SST system,  a number of the  basic qualitative results above are also found with simpler models\cite{lattice_model1,lattice_model2,Aleks_off_lattice,lattice_model_wang2,lattice_model_wang,Jacobs_protocol_design} and  are likely to generically hold  for a wider range of target structures\cite{PengYin3d,Peng_Yin_Dephts,Peng_yin_10nm,Peng_yin_tubes,Peng_yin_isothermal,Peng_yin_design_space,Peng_yin_reconfigurations} that use the addressable SST assembly method.

There are a number of possible extensions of our calculations.  A particular important direction will be to perform simulations in an NVT ensemble which would allow for a more careful comparison to collective properties (such as yield) that are measured in experiments, in particular the role of monomer starvation in reducing the yield at higher cooling rates could be quantitatively elucidated.  Extending our kinetic model to consider this situation is in principle straightforward. Future tests to our approach may include a careful compassion between the distribution of structure sizes observed in the experimental samples and that predicted by the kinetic model for a similar background concentration of monomers.

Making different shapes, as done in Ref.~\onlinecite{PengYin2d} by leaving out certain tiles, would also be straightforward to investigate, and it would be interesting to see how the nucleation pathway might change for a more complex shape.  Extending the method to consider SST structures in 3 dimensions is also feasible, but would of course require the detailed thermodynamics of the different local environments relevant to this new geometry to be characterized.

In this paper we used an average-base model that enables us to look at the generic behaviour of these SST systems. However, it should be relatively uncomplicated to extend the model to investigate a specific set of sequences by using the NN model to estimate how the relevant $\Delta G^\plimsoll$ would differ from that calculated for the average-base model.  The model could then be used to study how sequence might affect the nucleation pathway - one might imagine that nucleation is more likely to occur in regions with higher GC content - and how it could be used to optimise the assembly of particular structures.

Our methodology, where oxDNA is used to parameterise the rates for a limited set of geometries, which are then fed into a KMC scheme, can also be extended to other DNA tile systems such as 2D double-crossover tiles\cite{winfree1998design} which are widely used for studying molecular computation.

Lastly, we point out that our modelling approach is based upon the assumption that interactions relevant for tile-incorporation are typically local, and to a good approximation only involve components in the local neighbourhood of the binding site; as a consequence our modelling strategy may not be adequate to address self-assembling systems where relevant interactions are typically long ranged. In particular, the description of DNA origamis, in which strand incorporation typically involves bringing into close proximity distant sections of the circular scaffold strand (and is further complicated by other strands already incorporated in the structure\cite{dannenberg2015modelling}) is at the present beyond the scope of our multi-scale modelling approach.

\section*{supplementary material}

See supplementary material for further details on the simulations performed with oxDNA (S.I and S.II), misbond formation calculations (S.III and S.V), the structure of the critical nucleus at $T_c=51.64^\circ\text{C}$ (S.IV), the full set of 32 local geometries and respective free energy profiles (S.VI), an example of a bridging tile (S.VII), the tests performed to gauge the effect of detailed balance deviations (S.VIII and S.IX), and additional details on the experimental system considered to parametrise $k_+^0$  (S.X).

\begin{acknowledgements}

The authors wish to thank the EPSRC for financial support (EP/I001352/1 and EP/J019445/1). PF thanks FCT-Portugal for funding through grant SFRH/BD/94405/2013. TEO is supported by a Royal Society University Research Fellowship. We acknowledge the use of the University of Oxford Advanced Research Computing (ARC) facility in carrying out some of this work (doi:10.5281/zenodo.22558).

\end{acknowledgements} 



\bibstyle{aip}

\end{document}



\title{Supplementary Materials for ``Multi-scale coarse-graining for the study of assembly pathways in DNA-brick self assembly''}

\author{Pedro Fonseca}
\email{pedro.fonseca@physics.ox.ac.uk}
\affiliation{Rudolf Peierls Centre for Theoretical Physics, University of Oxford, 1 Keble Road, Oxford, OX1 3NP, United Kingdom}

\author{Flavio Romano}%
\affiliation{ Dipartimento di Scienze Molecolari e Nanosistemi,
Universit\'a Ca' Foscari, Via Torino 155, 30172 Venezia Mestre, Italy.}

\author{John S. Schreck}%
\affiliation{Department of Chemical Engineering, Columbia University, 500 W 120th St, New York, NY 10027, USA}

\author{Thomas E. Ouldridge}%
\affiliation{ Department of Bioengineering, Imperial College London, London, SW7 2AZ, United Kingdom}

\author{Jonathan P. K. Doye}
\affiliation{Physical and Theoretical Chemistry Laboratory, Department of Chemistry, University of Oxford, South Parks Road, Oxford, OX1 3QZ, United Kingdom}

\author{Ard A. Louis}
\affiliation{Rudolf Peierls Centre for Theoretical Physics, University of Oxford, 1 Keble Road, Oxford, OX1 3NP, United Kingdom}

\date{\today}
\maketitle

\onecolumngrid

\begin{figure}[ht]
    \centering
    \includegraphics[width=0.8\textwidth]{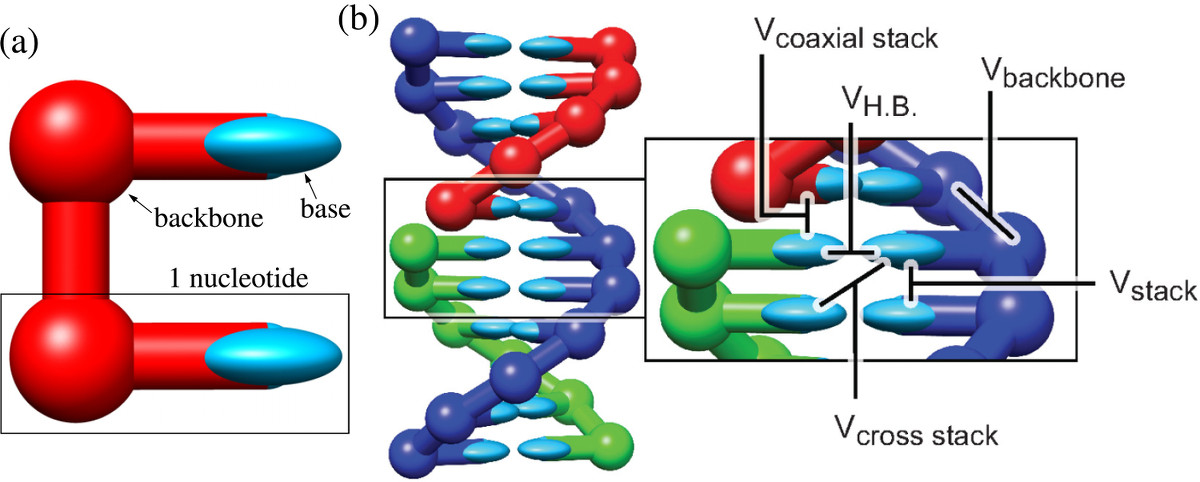}
    \caption{Representation of the oxDNA model. The backbone sites are represented as spheres; the bases are represented by ellipsoids emphasizing the orientation dependence of the interactions. (a) Representation of rigid nucleotides. (b) Representation of three hybridized stands. The pairwise interactions of the model are depicted in the inset. All nucleotides also interact through repulsive excluded volume interactions. Reproduced with permission of J. Schreck \textit{et al.}~\cite{Schreck_oxDNA_pic}}\label{fig:oxDNA}
\end{figure}

\section{oxDNA model}\label{oxDNA}

oxDNA~\cite{tom_model_jcp,oxdna_review} is a top-down model developed to study self-assembly processes associated with DNA nanotechnology. oxDNA makes use of three coarse-grained units per nucleotide plus the full orientation of the base to account for angle-dependent interactions. In Fig.~S\ref{fig:oxDNA} we show a schematic representation of the potentials: pairwise interactions between the nucleotides represent base-pairing between complementary Watson-Crick bases, stacking, coaxial stacking, and chain connectivity; in addition, all nucleotides also interact through repulsive excluded volume. oxDNA potentials were parametrised to reproduce the hybridisation transitions of duplexes as predicted by Santa Lucia's two-state nearest-neighbour (NN) model~\cite{SantaLucia1998,SantaLucia2004}, and some fundamental structural and mechanical properties of single-stranded DNA and duplex DNA. Since its conception, oxDNA has successfully predicted functional properties of many biophysical and nanotechnological systems beyond its original parametrisation. An extensive overview of the model can be found in references ~\cite{tom_model_jcp,oxdna_review}, and the source code and full list of publications can be accessed in the website \href{https://dna.physics.ox.ac.uk/index.php/Main_Page}{dna.physics.ox.ac.uk}.


\section{Simulation Methods}

\subsection{Virtual Move Monte Carlo (VMMC)}

Standard Monte Carlo (MC) methods, in which sequential one-particle moves are attempted, can be very inefficient to sample dilute systems with strong short-ranged interactions. This is the case for the systems studied throughout this work where moves in which a single nucleotide is displaced (the nucleotide is the basic constituent of DNA within oxDNA) suffer from low acceptance rates, resulting in strongly suppressed collective motion of strands and slow sampling of diffusion. Simulations can be made more efficient by using the VMMC algorithm proposed by Whitelam and Geissler~\cite{Whitelam_VMMC}, which allows for collective moves involving clusters of nucleotides. 

To generate a trial move the algorithm first selects a single nucleotide, and attempts a random seed move according to standard Metropolis rules. Additional nucleotides in the neighbourhood of the selected nucleotide are then added, with probabilities that depend on the resultant energy changes, forming a co-moving cluster. Trial moves are then accepted (or rejected) with a probability that ensures correct sampling from the canonic ensemble. In the simulations performed through this work, seed moves were: 

\begin{itemize}
\item Rotation of a nucleotide about its backbone site, with the axis chosen uniformly on the unit sphere and the angle drawn from a normal distribution with a mean of zero and a standard deviation of 0.22 radians.
\item Translation of a nucleotide, where the displacement along each Cartesian axis is drawn from a normal distribution with a mean zero and a standard deviation of 1.28 $\textnormal{\AA} $.
\end{itemize}





\subsection{Single Histogram Reweighting}

In this work, we  use oxDNA extensively to calculate tile-association free energies at $T=50^\circ  \textnormal{C}$, in the temperature range we expected the assembly transition to be (the experimental peak of the rate-of-assembly curve is at $T=53^\circ  \textnormal{C}$). To avoid having to repeat simulations for temperatures other than $T=50^\circ  \textnormal{C}$, we follow earlier work on oxDNA~\cite{vsulc2012sequence} and use the single histogram reweighting method,  first introduced by Ferrenberg and Swensden~\cite{SHR_method}, to extrapolate our results to other temperatures.

To use this method, the states of the system sampled in the simulations are grouped according to their energy $E$, and value of some observable of interest $A$, and the distribution $p(A,E;T_0)$ is produced during the run (all runs were performed at $T_0=50^\circ  \textnormal{C}$). From the distribution $p(A,E;T_0)$, we can derive the density of states $p(A,E;T_0)\propto \Omega (A,E)e^{-\beta_0 E}$, where $\beta_0=1/k_BT_0$. The density of states can then be used to estimate values of $A$ at a different temperature $T$:  

\begin{equation}
\langle A(T) \rangle = \frac{\int \int A\Omega (A,E)e^{-\beta E}dEdA}{\int \int \Omega(A,E)e^{-\beta E}dEdA},
\end{equation}\label{eq:eq2}

where $\beta=1/k_BT$. Eq. (2) can be rewritten as: 

\begin{equation}
\langle A(T) \rangle = \frac{\int \int Ap(A,E;T_0)e^{-(\beta_0-\beta) E}dEdA}{\int \int p(A,E;T_0)e^{-(\beta_0-\beta) E}dEdA},
\end{equation}\label{eq:eq3}


\section{Misbond Formation Calculations}

\begin{figure}[h]
    \centering
    \includegraphics[width=0.45\textwidth]{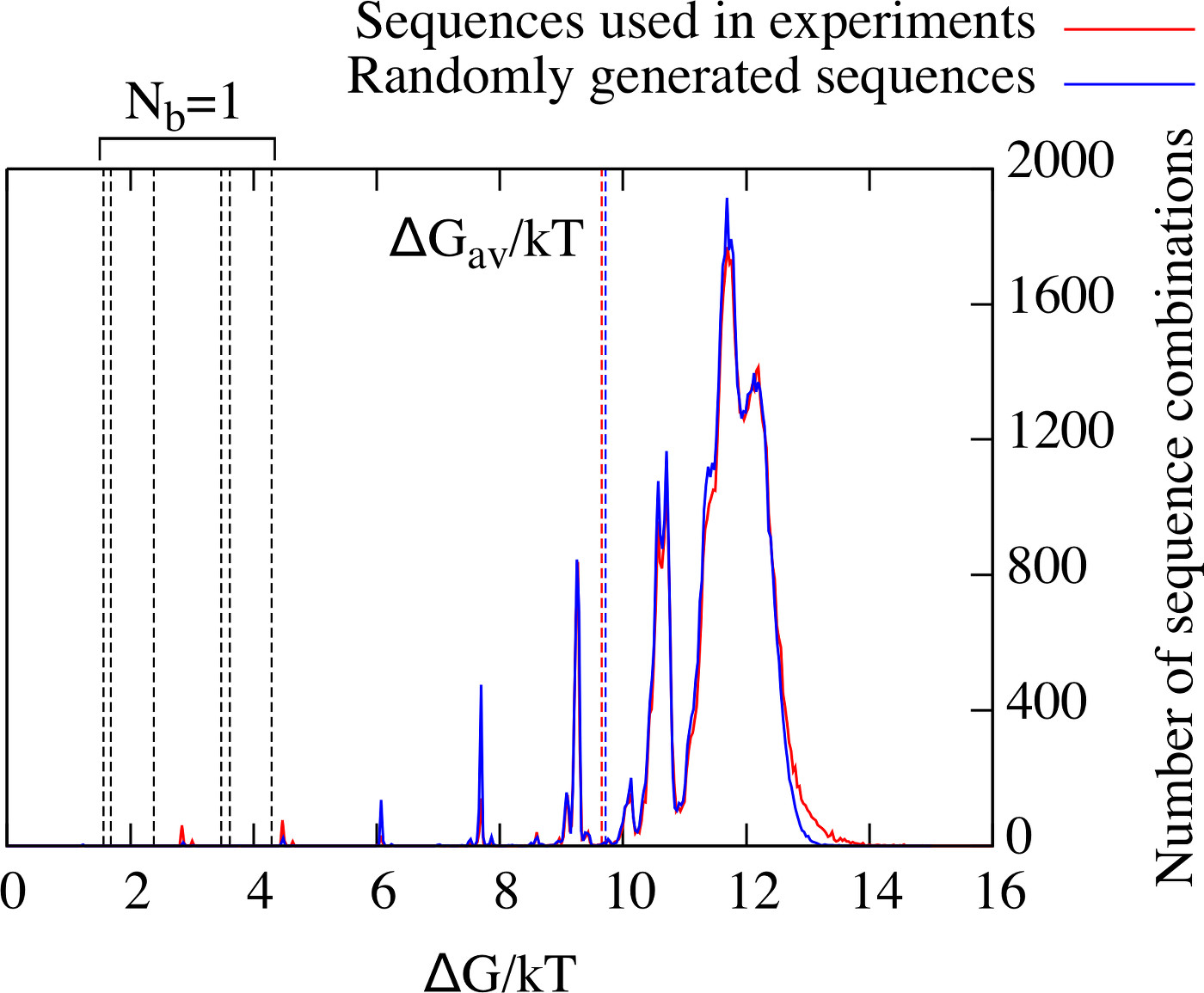}
    \caption{Effective free energies of interaction determined for pairs of sequences (each 42-base long) that were not designed to interact, estimated using the NN model~\cite{SantaLucia1998}, at 100nM per strand species and $T = 50^\circ\text{C}$. Red curve: calculation using sequences used in experiments~\cite{PengYin2d}. Blue curve: calculation using randomly generated sequences. Dashed coloured lines: contact free-energies ($\Delta G_{\textnormal{av}}/kT$) correspondent to the average probability of bonding between strands (taken over all combinations of sequences). Also shown for comparison (dashed black lines) are the free energies of tile association determined using oxDNA for binding sites with a single domain available for binding ($N_b=1$), also at $100\, \textnormal{nM}$ per strand species and $T = 50^\circ\text{C}$.}\label{fig:dg_misbonds}
\end{figure}


In this section we estimate an upper bound on the equilibrium probability of misbonding at binding sites at the interface of the assembly. We also estimate the effect of misbonds on assembly dynamics. In the following sections we will use the term ``binding site type'' to differentiate between binding sites that have a different set of first neighbours.

We start our analysis by observing that for all binding site types, the cumulative length of binding domains is at most 42-bases long (see S.VI, environments  27-32).  For most binding site types however, the cumulative length of binding domains is less than 42 bases (see S.VI, environments 1-26), but for simplicity we assume it to be 42 bases for all binding site types. Note that by using this simplification, we overestimate the stability of misbonds and thus, also the probability of misbonding at equilibrium. 

We further simplify our analysis by assuming that, instead of being distributed through multiple neighbouring tiles, binding domains can be treated as a single continuous segment with 42 bases. Again, we note that this simplification overestimates the probability of misbonding at equilibrium - as can be seen in the free-energy profiles of tile-association (see S.VI, environments 7-32) there is a significant entropic penalty associated with initiating a second or subsequent domains, which is not accounted for if binding domains are considered to be concatenated in a single-stranded segment. We emphasise that our goal is to estimate an upper bound on the equilibrium probability of misbonding rather than obtaining a precise value.






\subsection{Calculation of misbond free energies}

We generate a representative set of 42-base sequences for the sites by using the sequences of the 334 tiles themselves. The challenge then reduces to estimating the binding free energy of pairs of sequences (each 42-base long) from the ensemble used in experiments~\cite{PengYin2d}. We exclude from the calculations sequences that share a complementary domain designed to hybridise in the assembled structure, which leaves a total of $334\times 333/2 - 616=54\, 995$ different combinations of 2 sequences. For each pair of sequences, we chose one ($A$) to represent a binding site in the assembly, and the other ($s$) to represent an incoming strand that may incorrectly form bonds with the binding site. We then use use a simple algorithm that finds complementary sections between both sequences that are at least 2 bases long. For each of the complementary sections found, and all pairs of such sections, we calculate the free energy of the base-pairing configuration using the average-base parameterisation of the NN model~\cite{SantaLucia2004}, together with internal loop parameters and asymmetry corrections~\cite{SantaLucia2004,SantaLucia2004}. To each different complementary section between $A$ and $s$ (or two simultaneously), we add a label $\eta$ to which we call the ``contact type" between $A,s$. 



For each contact type $\eta$ between a pair of sequences $A,s$ the equilibrium probability of contact type $\eta$ being formed with relation to the reference state $\O$ (where no contacts are formed) is given by: 
 \begin{equation}\label{eq:eq5}
 p_\eta(A,s)/p_{\O}(A,s)=  e^{-\Delta \beta G^{\plimsoll}_{\eta}(A,s)} \frac{[s]}{ 1\textnormal{M}} = \kappa_{\eta}(A,s).
  \end{equation}
Using $ p_{\O}(A,s)+\sum_{\eta}p_\eta(A,s)=1$, we determine the probability of \textit{any} contact being formed between the pair $A,s$, which we denote by $p(A,S)\equiv \sum_{\eta}p_\eta (A,s)$:
\begin{equation}\label{eq:eq6}
p(A,s)=\frac{\sum_{\eta} \kappa_{\eta}(A,s)}{1+\sum_{\eta} \kappa_{\eta}(A,s)}. \end{equation}
Finally, we determine an overall free energy of interaction for the pair $A,s$ at the concentration $[s]$, $\Delta G^\plimsoll({A,s})$, by solving the equation: 
\begin{equation}\label{eq:eq8}
p(A,s)/p_{\O}(A,s)= \kappa (A,s) = e^{-\beta \Delta G^\plimsoll(A,s)} [s] /(1\textnormal{M}).
 \end{equation}
 

In Fig.~S\ref{fig:dg_misbonds} we show the resulting distribution of effective free energies of interaction ($\Delta G^{\plimsoll}(A,s)$) obtained by repeating the calculation for the different $54995$ combinations of two sequences from the set used in the experiments (red curve). Also shown in the figure (blue curve) is the distribution of effective free energies of interaction determined using a similar calculation, but this time using a batch of strands whose sequences were generated at random (with the constraint of observing the complementary pattern needed for the target structure to assemble).



\subsection{Estimation of the effect of misbonds on equilibrium and dynamic properties of assembly}

To estimate the effect of misbonds on assembly at equilibrium, we first consider the average probability at equilibrium that a given strand with an incorrect sequence binds to a binding site of the assembly. In this calculation, we take the average over binding probabilities for all combinations of representative binding site and tile sequences:  $\bar{p}=\sum_{A,S} p(A,s)/M$, with $M=54995$ being the total number of different pairs of sequences. Next, we calculate the interaction free energy $\Delta G_{\textnormal{av}}^{\plimsoll}$ corresponding to $\bar{p}$ by solving the equation:


 \begin{equation}\label{eq:eq8a}
\frac{\bar{p}}{(1-\bar{p})}= \kappa_{\textnormal{av}}=e^{-\Delta G_{\textnormal{av}}^{\plimsoll}/RT}[s] /(1\textnormal{M}).
 \end{equation}

In Fig.~S\ref{fig:dg_misbonds} we show (dashed coloured lines) the free energies $\Delta G_{\textnormal{av}}^{\plimsoll}$ determined using sequences taken from the experimental batch (red) and sequences generated randomly (blue).


\subsubsection{Misbonding model}

We are now in a position to incorporate this overestimate of typical mismatch binding propensity into the self-assembly model presented in the main text. 
We assume that for each binding site, there are 333 incorrect strands, all present at concentration $[s]$, which all bind with a bimolecular rate constant $k_+^0$, yielding an overall misbonding rate
  \begin{equation}\label{eq:commulative_rate}
k^{\mathrm{mm}}_+=333\times k_+^0 [s],
 \end{equation}
 and dissociate at a rate
\begin{equation}\label{eq:disscommulative_rate}
k^{\mathrm {mm}}_{-}=k_{+}^{0} e^{\beta \Delta G_{\textnormal{av}}^{\plimsoll }}\times (1\textnormal{M}).  
\end{equation}
This is the simplest kinetic approach that captures the overall tendency of misbonds to form in equilibrium, whilst averaging over the sequence details.

 
 
 
\subsubsection{Overall effect of misbonding on assembly at equilibrium} 

Binding sites in the assembly can be either unoccupied or occupied by either a correctly bonded tile or a strand with an incorrect sequence. To estimate the 
relative probabilities in equilibrium we use the rates in Eqs.~\ref{eq:commulative_rate}, \ref{eq:disscommulative_rate} and Eqs. 1 and 2 of the main text. 
For the relative probabilities of the misbonded and unoccupied states, we have
\begin{equation}\label{eq:balance_mis}
p_{\mathrm{misbond}}/p(\O '' )={k^{\mathrm {mm}_+}}/{k^{\mathrm {mm}_-}}, 
 \end{equation}
For the correct tile, from Eq.~2 of the main text, we have
\begin{equation}\label{eq:balance_cor}
p_{(\textnormal{correct})}/p(\O '')= \exp\left({-\beta \Delta G^{\plimsoll}(\vec{x},\vec{y})}\right)[s]/1\textnormal{M},  
 \end{equation}
which depends on the nature of the states $\vec{x}$ and $\vec{y}$ with empty and full sites, respectively. 

\begin{figure}[h]
    \centering
    \includegraphics[width=0.45\textwidth]{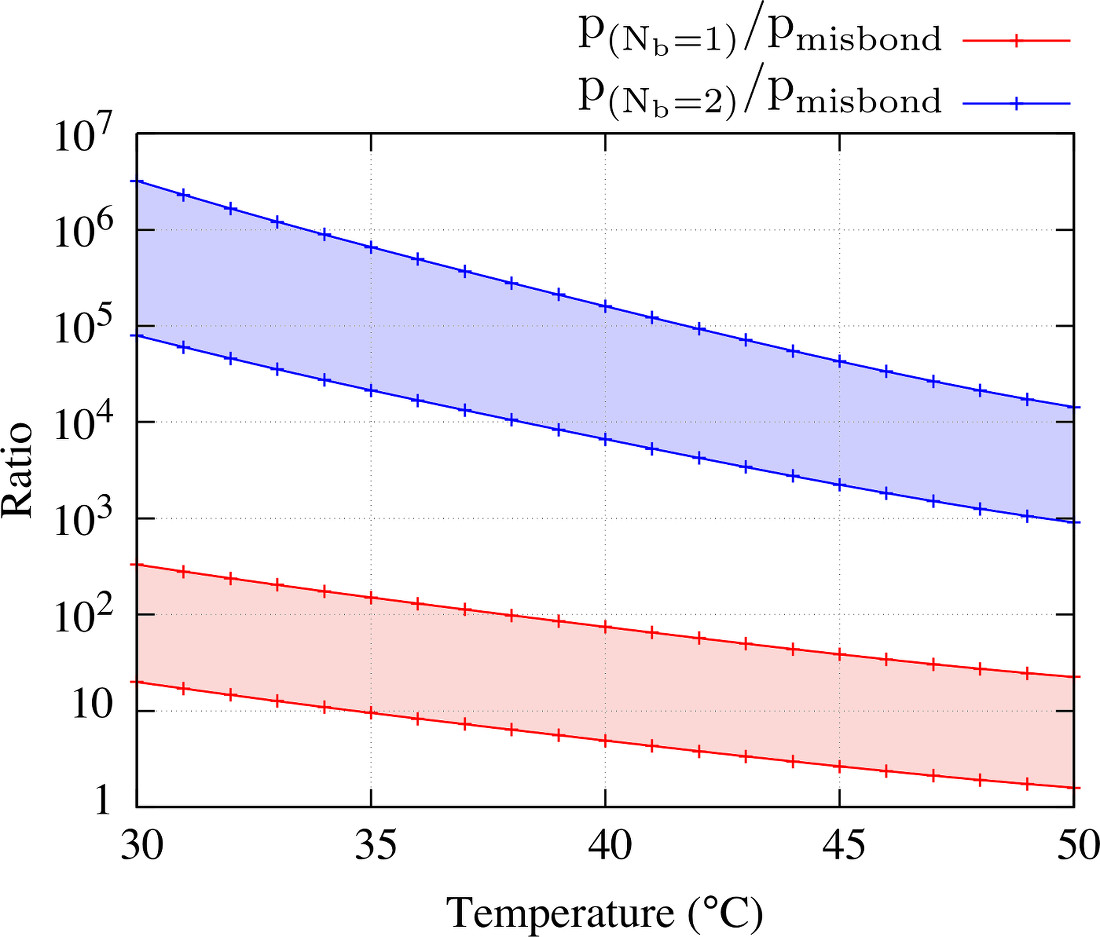}
    \caption{Ratio between the equilibrium probability of bonding for correctly bonded tiles with with one (${p}_{(\textnormal{N}_b=1)}$) or two (${p}_{(\textnormal{N}_b=2)}$) binding domains, and the equilibrium probability of misbonding (${p}_{\text{misbond}}$) as a function of temperature.  The solid curves limiting the coloured areas correspond to bonding probabilities for the least strongly bonded (bottom curve) and most strongly bonded (top curve) \tom{correct} tiles.}\label{fig:pmisbond}
\end{figure}

In Fig.~S\ref{fig:pmisbond} we show the ratio of the equilibrium binding probabilities for correct bonding and misbonding, incorporating the variation over site type into the estimate of correct bonding.

\subsubsection{Effect of misbonding on assembly dynamics} 
 
 To estimate the effect of misbonding on dynamics, we use a simplified variant of our assembly model, where tiles with the correct sequence can associate the assembly with rate $k_+^0 [s]$ but are not allowed to leave the assembly once bound (which is the typical behaviour we observe at lower temperatures ($T\lesssim 43^\circ \text{C}$)), and misbonding is described explicitly using the rates determined above (Eq.~\ref{eq:commulative_rate} and Eq.~\ref{eq:disscommulative_rate}).

\begin{figure}[h]
    \centering
    \includegraphics[width=0.45\textwidth]{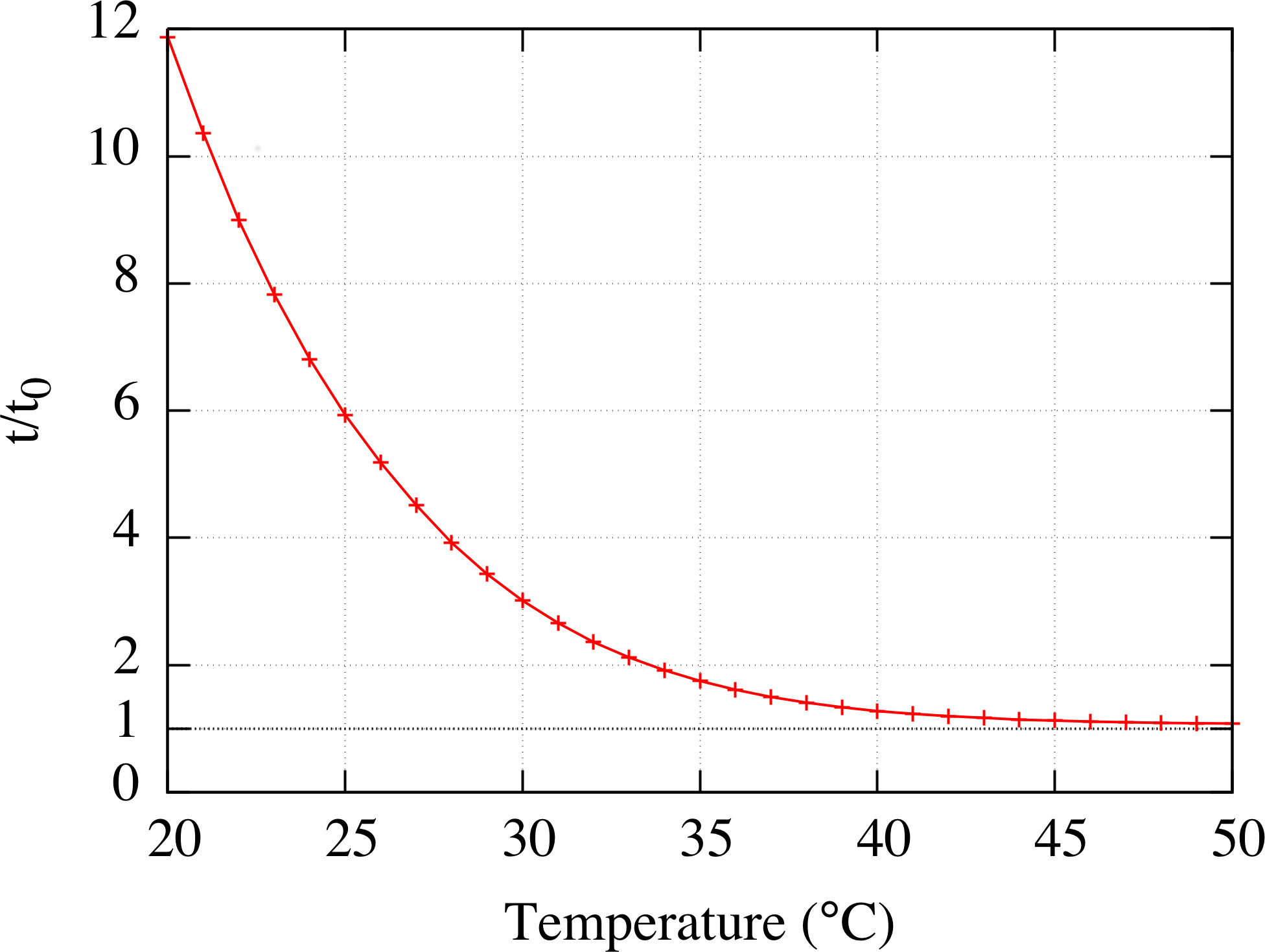}
    \caption{Ratio between average completion time when misbonds are explicitly described in the model ($t$), and average completion time when misbond are not allowed to occur ($t_0$).}\label{fig:tmisbond}
\end{figure}

 
 We perform our calculations using two sub-variants of the model. In the first we allow misbonds to occur (as described above) in addition to correct bonding; whereas in the second  sub-variant only correct bonding is allowed to occur.  We define $t$ as the average time needed for the assembly to reach full completion in the first sub-variant of the model, i.e. to reach a state where 334 tiles are correctly bonded; and define $t_0$ as the average completion time observed using the second sub-variant of the model. In Fig.~S\ref{fig:tmisbond} we show the ratio between average completion times observed in the two sets of simulations as a function of temperature. 
 

  \section{Critical nucleus structure}
 
\begin{figure}[h]
    \centering
    \includegraphics[width=0.45\textwidth]{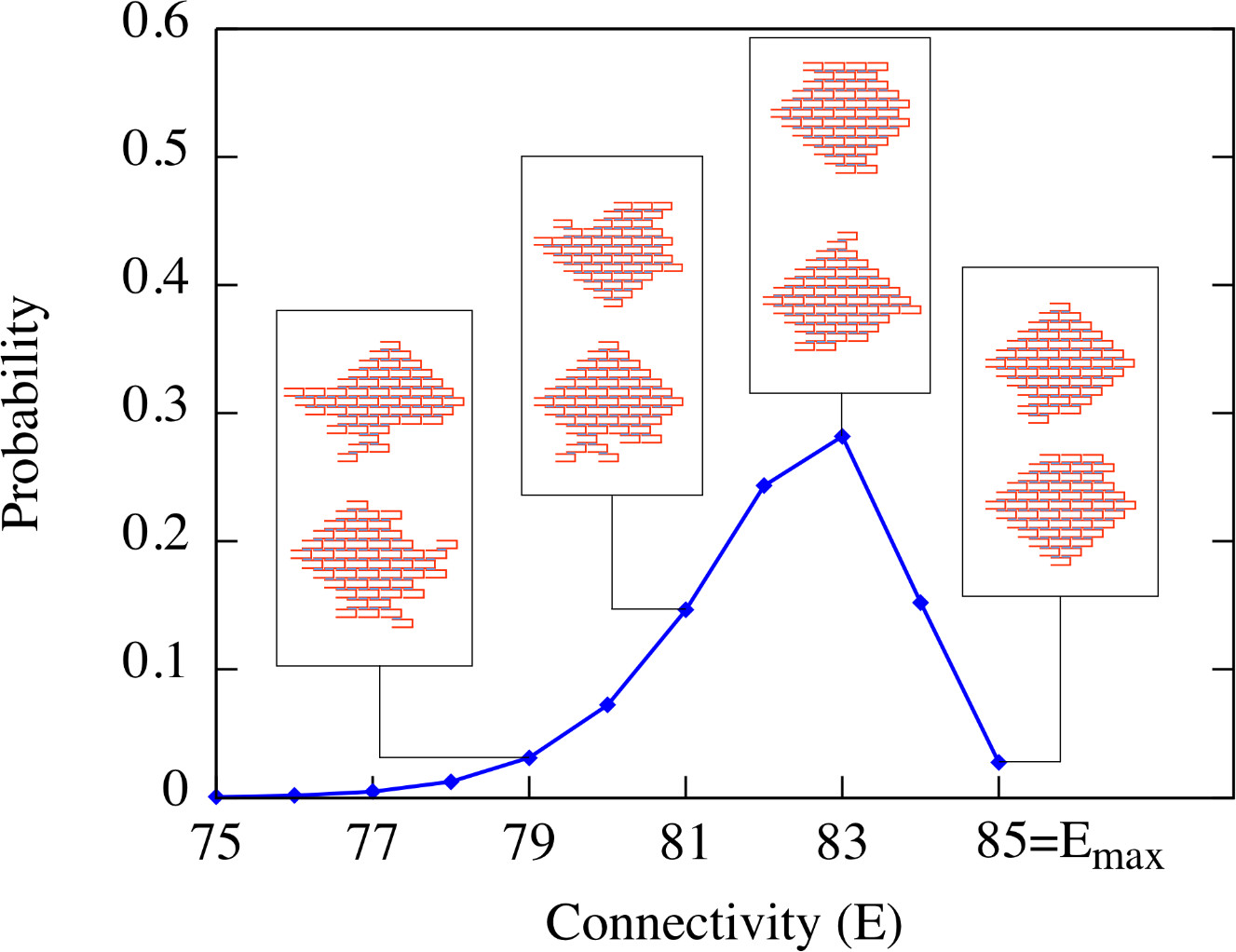}
    \caption{Probability distribution for the number of bonded domains (E) between tiles for structures with $N=50$ components (critical size,critical temperature - $T = 51.64^\circ\text{C}$). Above the curve are depicted examples of assembly states, representative of states  $(N^*=50,E)$ with different connectivities.}\label{fig:P_E_Tc}
\end{figure}

 As described in the main text, we calculated the probability of finding a critical nucleus with a certain number of bonded domains $E$. This is shown in Fig.~S\ref{fig:P_E_Tc} for the critical temperature $T = 51.64^\circ\text{C}$ where the critical nucleus size is $N^*=50$. While the structure with the maximum $E_{\text{max}}=85$ has the lowest enthalpy, it is not the most common value of $E$ because it also has a lower entropy.  Instead the peak in probability is around $E=83$.
 
 \section{Randomly generated sequences - probability distribution of the length of the longest complementary section between 2 sequences}
 
   In this section we determine the probability distribution for the length of the  longest complementary section between two randomly generated sequences, each 42-base-long. For this purpose, we use an algorithm that generates pairs of 42-base-long random sequences, checks for complementary sections between the two sequences, and stores the length (in number of bases) of the longest complementary section found. In Fig.~S\ref{fig:random_lengths} we show the resulting probability distribution, obtained by using the algorithm to produce and analyse $10^7$ pairs of random sequences.  

 \begin{figure}[h]
    \centering
    \includegraphics[width=0.45\textwidth]{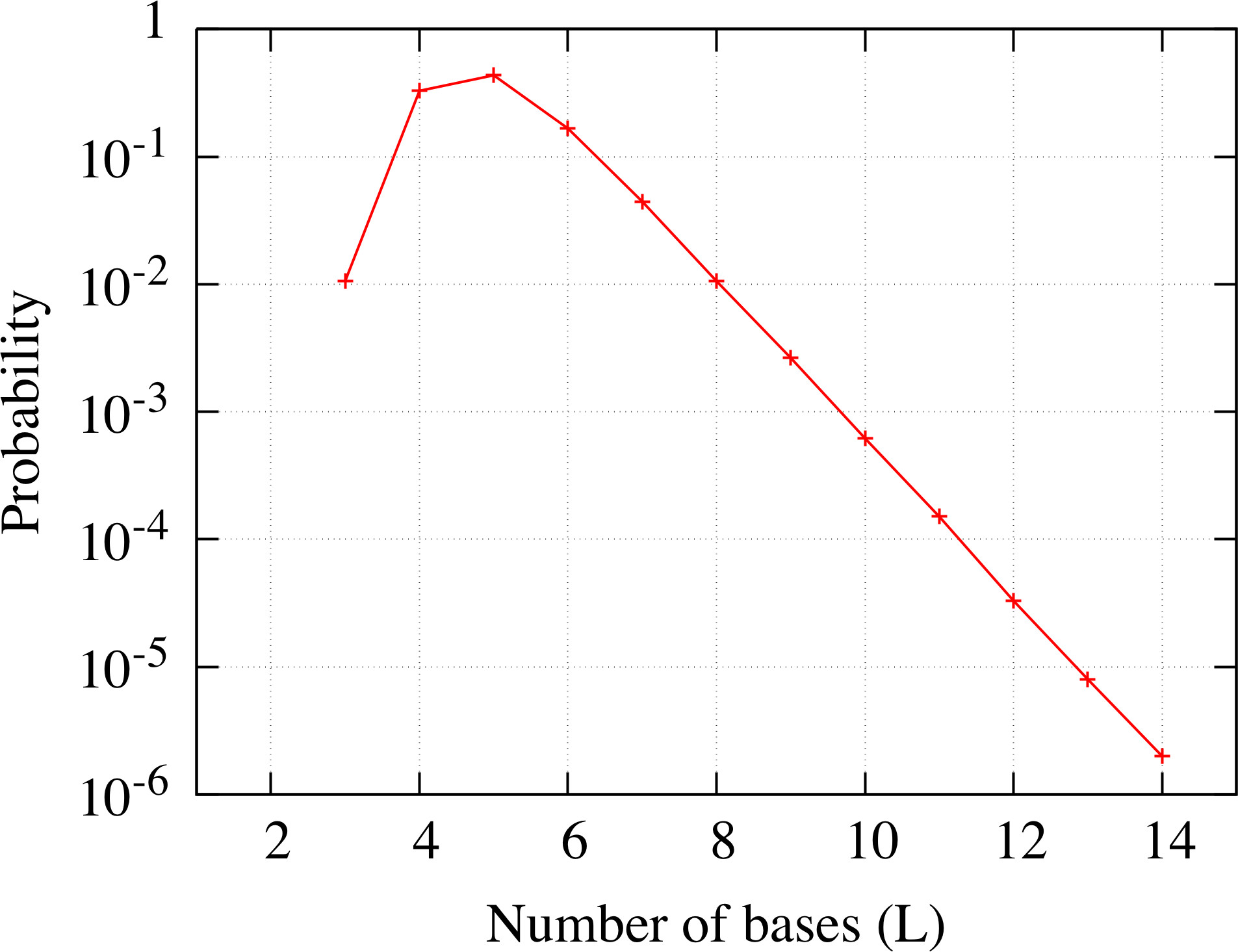}
    \caption{Probability distribution for the length ($L$) of the longest complementary section between two randomly generated (42-base-long) sequences.}\label{fig:random_lengths}
\end{figure}

\section{Free-energy profiles of tile-association}

In this section we detail the set of local environments used in the implementation of the kinetic model, and the respective free-energy profiles of tile-association calculated using oxDNA, at $T=50^\circ \textnormal{C}$ and $100\, \textnormal{nM}$ per tile species.  The numbers above each diagram (top left), are used to identify each local environment (also called environment index in the main text). In the schemes on the left, tiles coloured red represent the assembly fragment that contains the first neighbours of the binding site (which is colored gray). Rectangles coloured blue and grey represent duplex sections in the assembly fragment, and between assembly fragment and binding tile, respectively. The numbers in the schemes (coloured blue) represent the length (in nucleotides) of domains positioned above and bellow the numbers. In the plots on the right, $\textnormal{N}_{\textnormal{H-bonds}}$ denote the the number of base-pairs formed between tile and assembly fragment. 


\subsection{Local environments with a single binding domain}

\begin{figure}[h]
    \centering
    \includegraphics[width=0.90\textwidth]{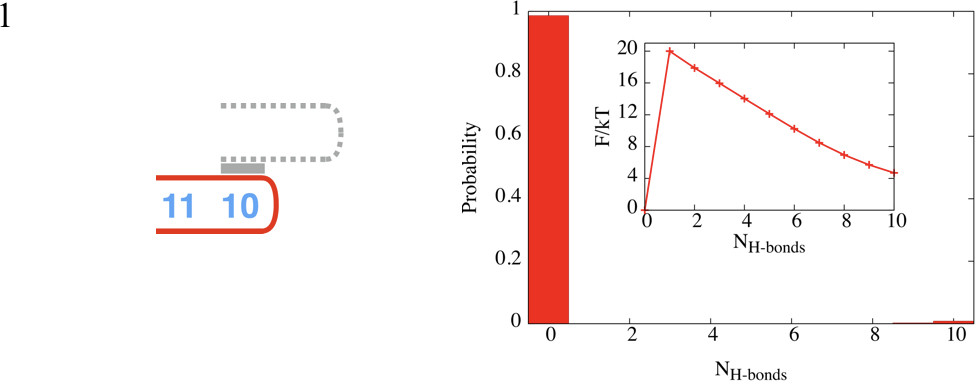}
\end{figure}

\begin{figure}[h]
    \centering
    \includegraphics[width=0.90\textwidth]{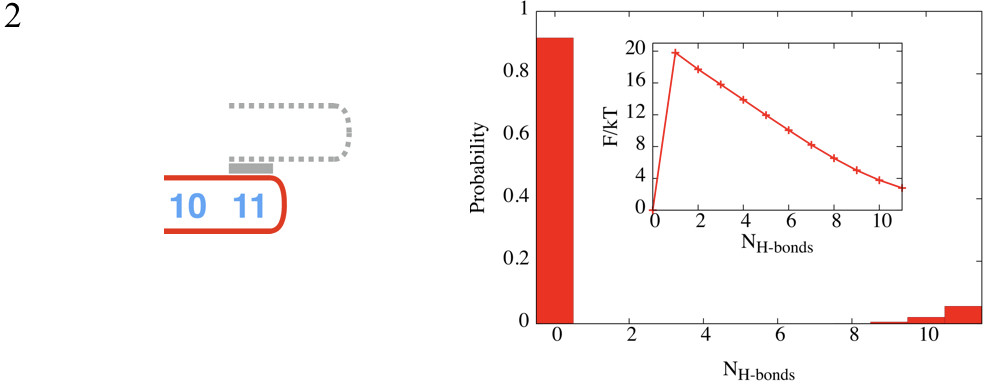}
\end{figure}

\newpage

\begin{figure}[h]
    \centering
    \includegraphics[width=0.90\textwidth]{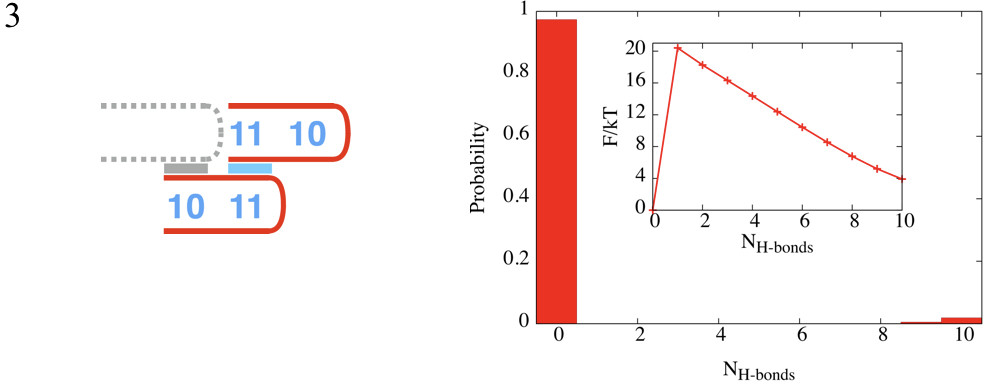}
\end{figure}

\begin{figure}[h]
    \centering
    \includegraphics[width=0.90\textwidth]{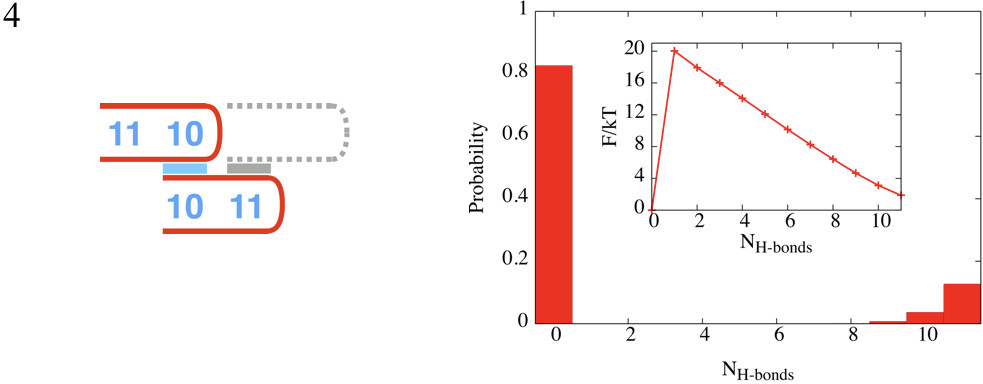}
\end{figure}

\begin{figure}[h]
    \centering
    \includegraphics[width=0.90\textwidth]{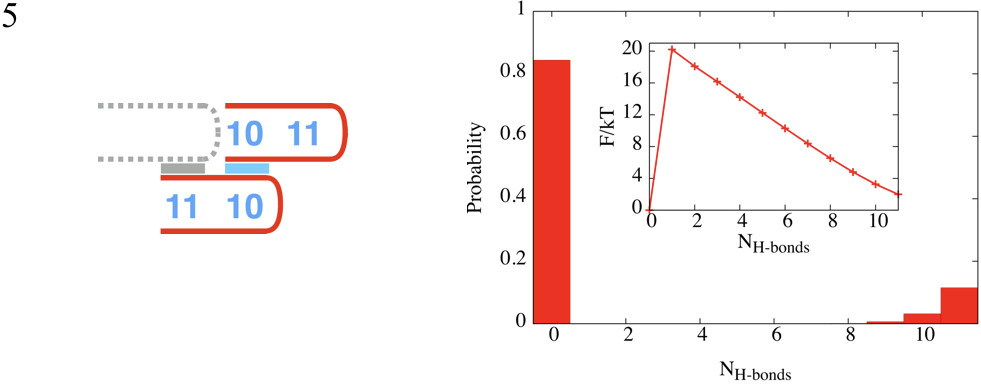}
\end{figure}

\newpage

\begin{figure}[h]
    \centering
    \includegraphics[width=0.90\textwidth]{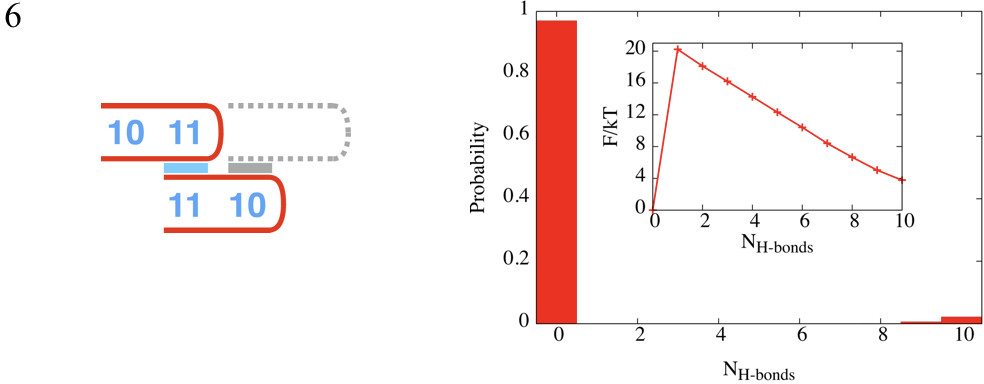}
\end{figure}

\subsection{Local environments with two binding domains}

\begin{figure}[h]
    \centering
    \includegraphics[width=0.90\textwidth]{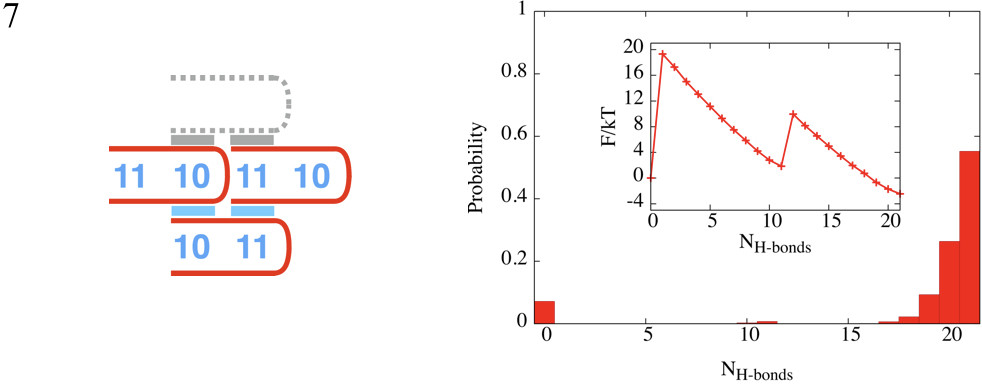}
\end{figure}

\begin{figure}[h]
    \centering
    \includegraphics[width=0.90\textwidth]{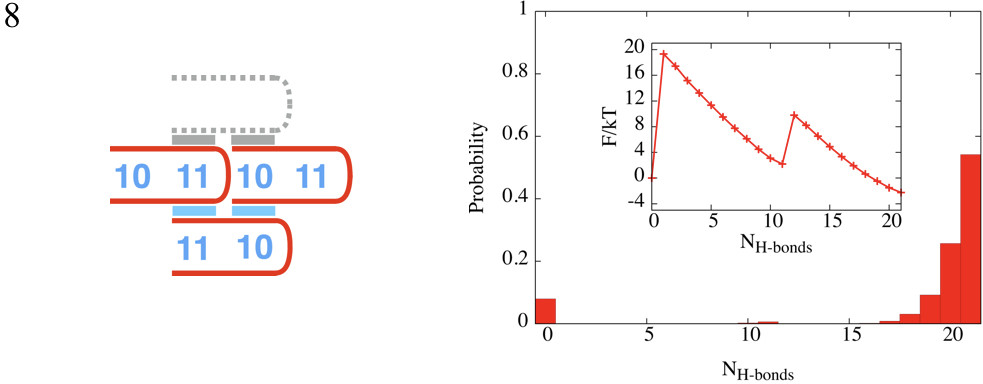}
\end{figure}

\newpage

\begin{figure}[h]
    \centering
    \includegraphics[width=0.90\textwidth]{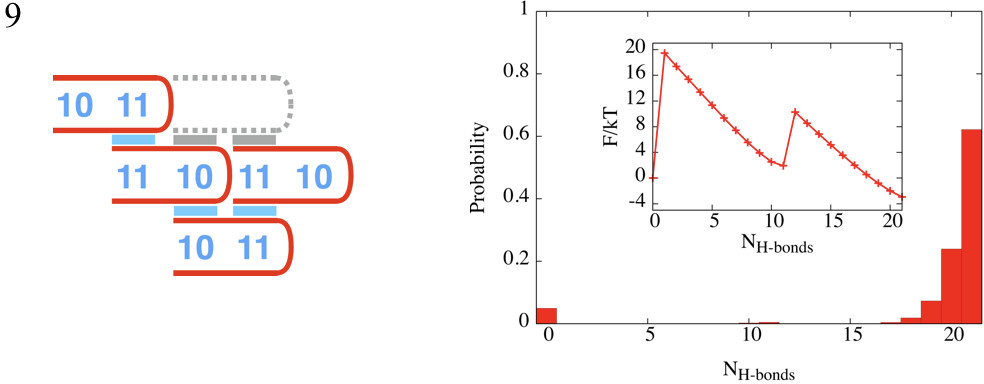}
\end{figure}

\begin{figure}[h]
    \centering
    \includegraphics[width=0.90\textwidth]{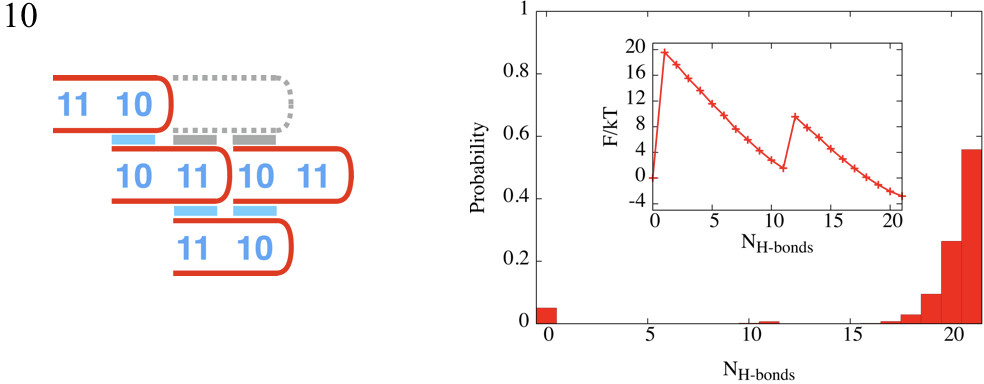}
\end{figure}

\begin{figure}[h]
    \centering
    \includegraphics[width=0.90\textwidth]{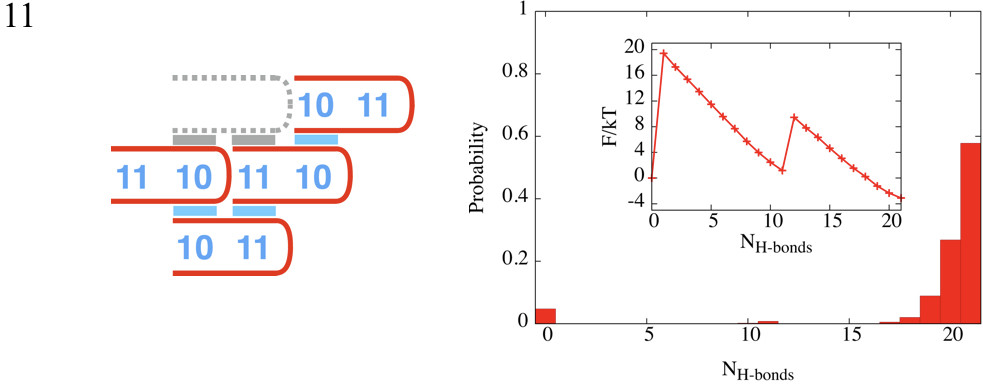}
\end{figure}

\newpage

\begin{figure}[h]
    \centering
    \includegraphics[width=0.90\textwidth]{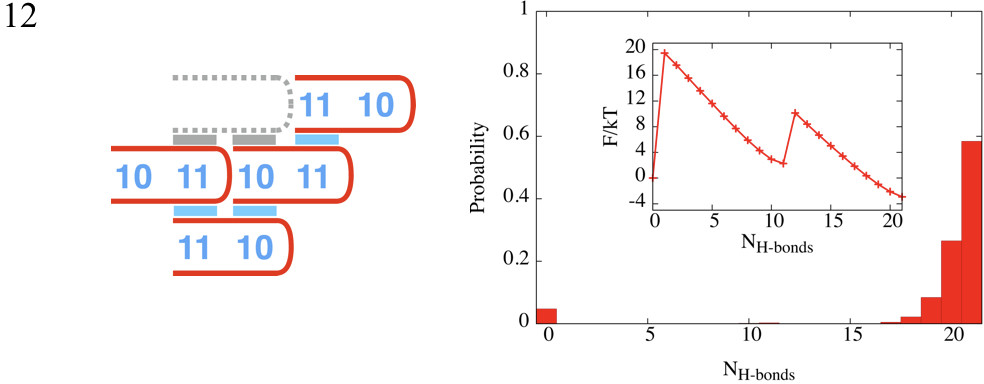}
\end{figure}

\begin{figure}[h]
    \centering
    \includegraphics[width=0.90\textwidth]{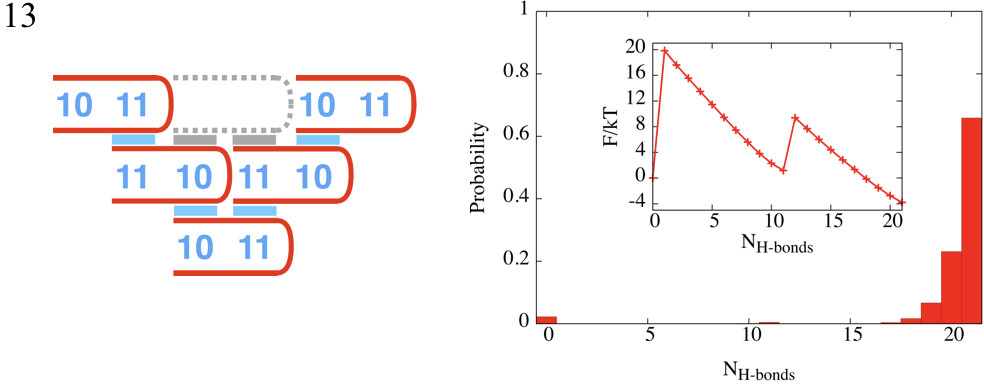}
\end{figure}

\begin{figure}[h]
    \centering
    \includegraphics[width=0.90\textwidth]{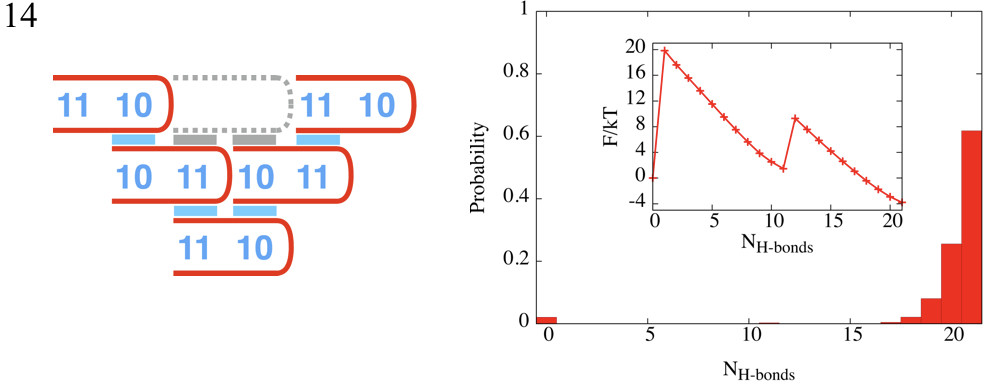}
\end{figure}

\newpage

\begin{figure}[h]
    \centering
    \includegraphics[width=0.90\textwidth]{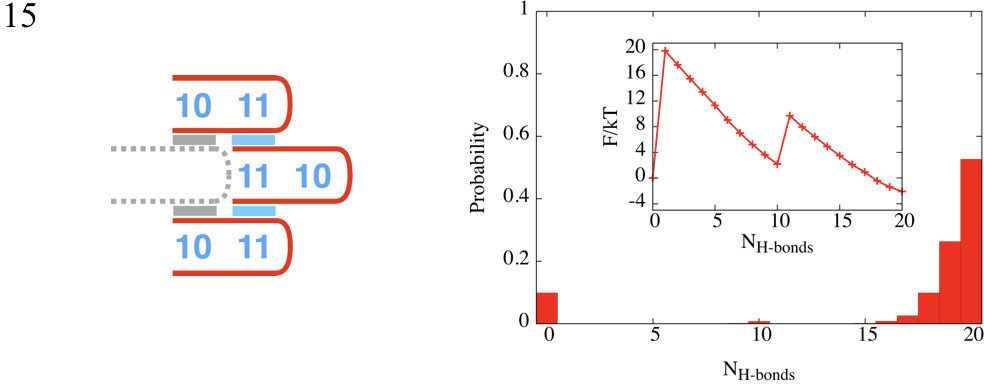}
\end{figure}

\begin{figure}[h]
    \centering
    \includegraphics[width=0.90\textwidth]{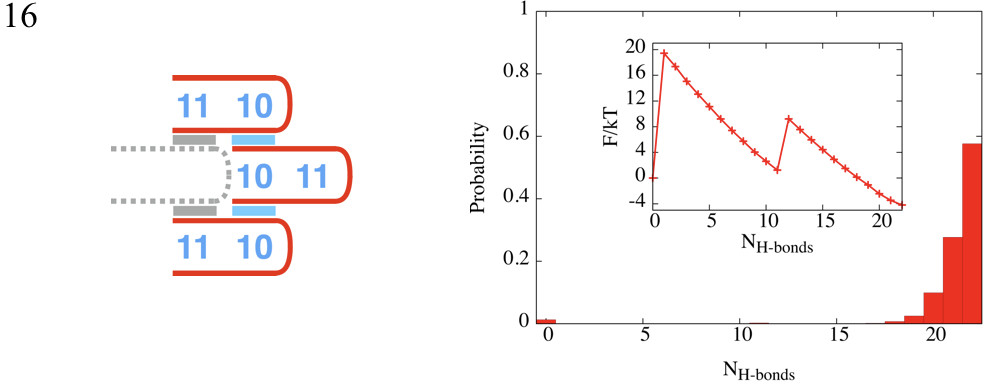}
\end{figure}

\begin{figure}[h]
    \centering
    \includegraphics[width=0.90\textwidth]{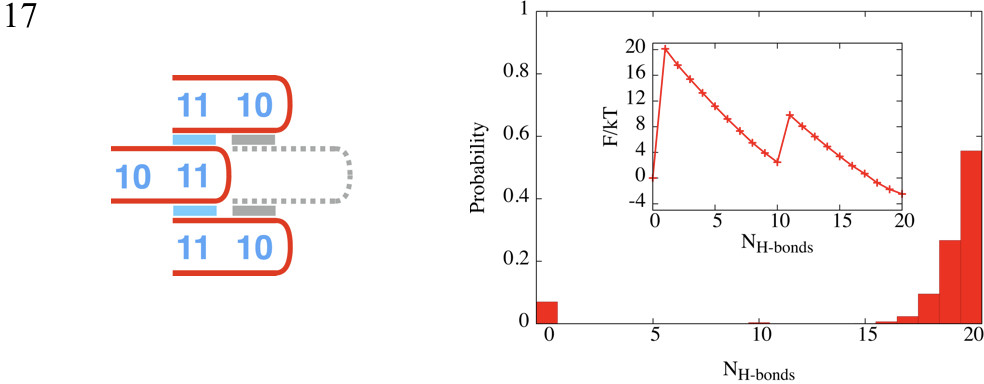}
\end{figure}

\newpage

\begin{figure}[h]
    \centering
    \includegraphics[width=0.90\textwidth]{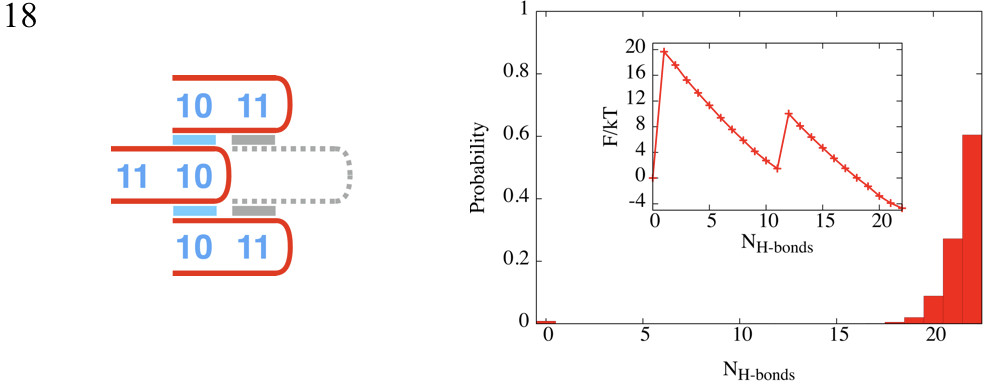}
\end{figure}

\subsection{Local environments with three binding domains}

\begin{figure}[h]
    \centering
    \includegraphics[width=0.90\textwidth]{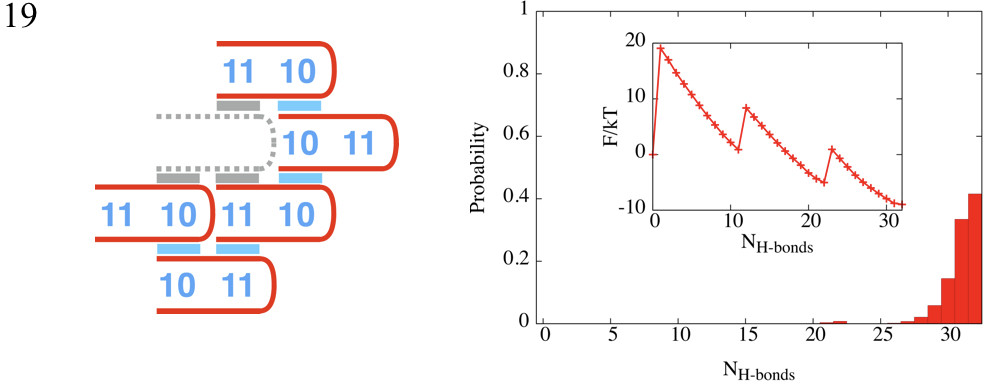}
\end{figure}

\begin{figure}[h]
    \centering
    \includegraphics[width=0.90\textwidth]{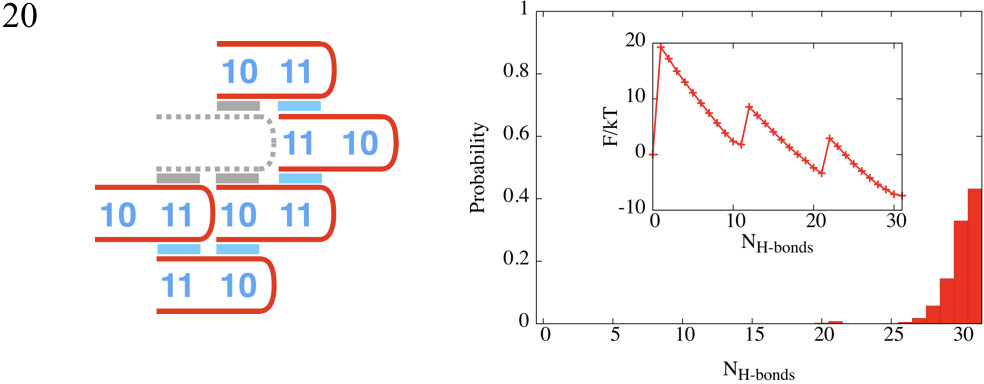}
\end{figure}

\newpage

\begin{figure}[h]
    \centering
    \includegraphics[width=0.90\textwidth]{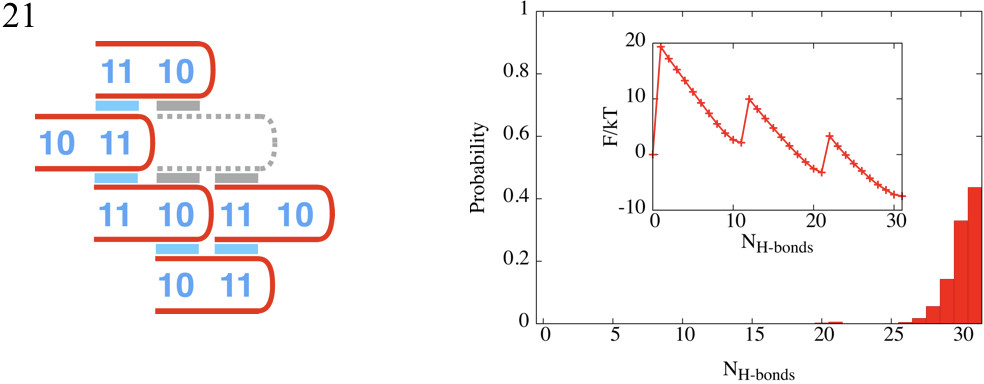}
\end{figure}

\begin{figure}[h]
    \centering
    \includegraphics[width=0.90\textwidth]{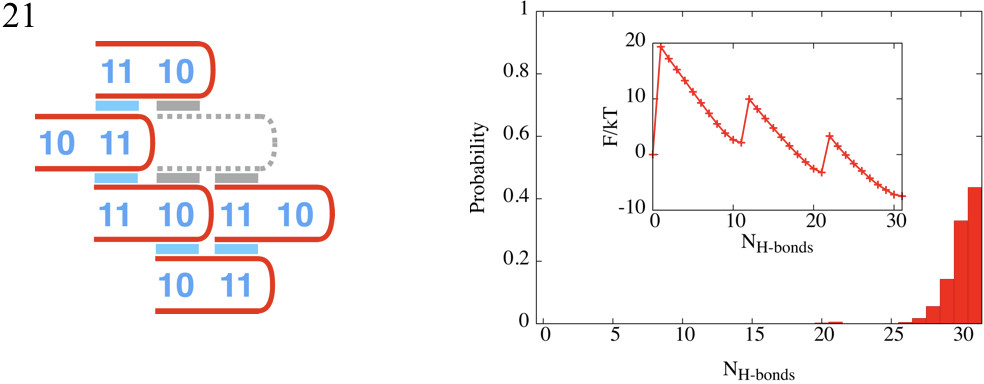}
\end{figure}

\begin{figure}[h]
    \centering
    \includegraphics[width=0.90\textwidth]{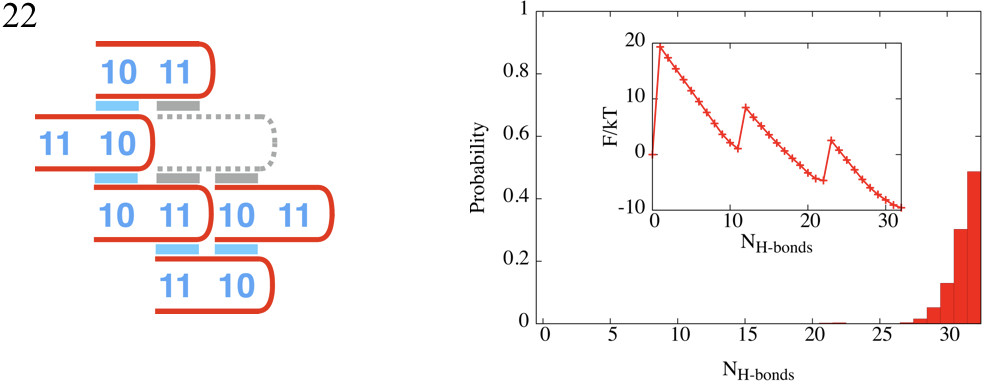}
\end{figure}

\newpage

\begin{figure}[h]
    \centering
    \includegraphics[width=0.90\textwidth]{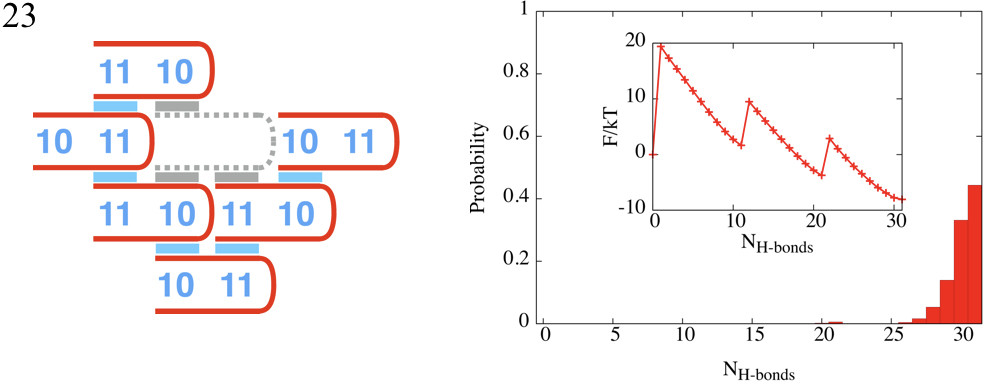}
\end{figure}

\begin{figure}[h]
    \centering
    \includegraphics[width=0.90\textwidth]{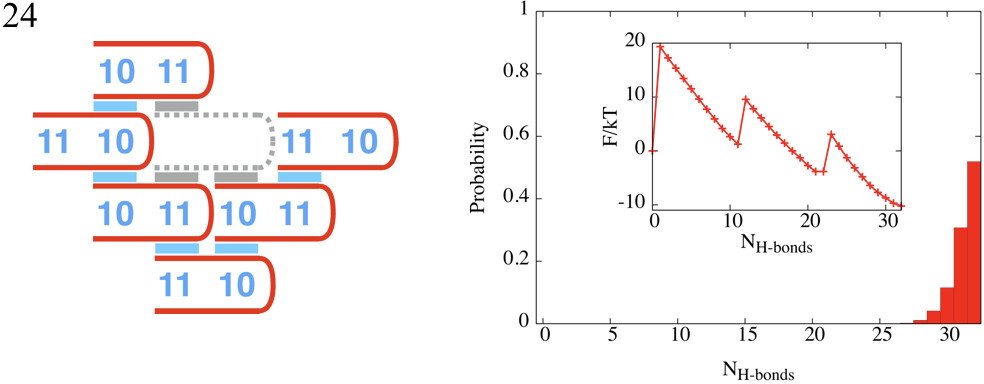}
\end{figure}

\begin{figure}[h]
    \centering
    \includegraphics[width=0.90\textwidth]{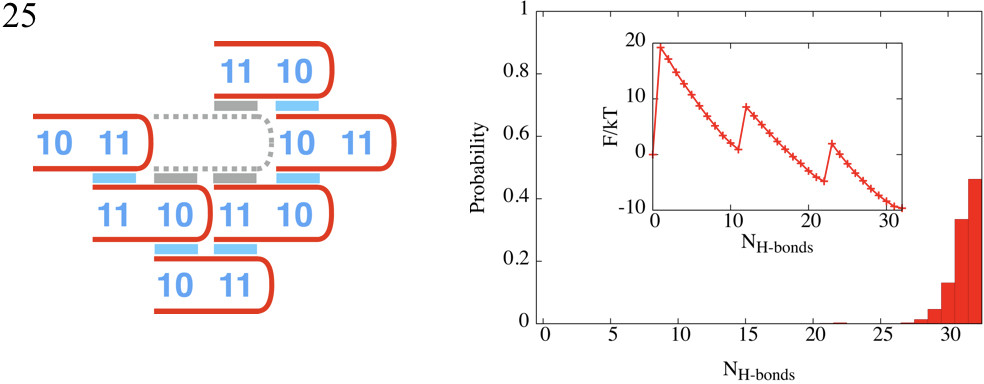}
\end{figure}

\newpage

\begin{figure}[h]
    \centering
    \includegraphics[width=0.90\textwidth]{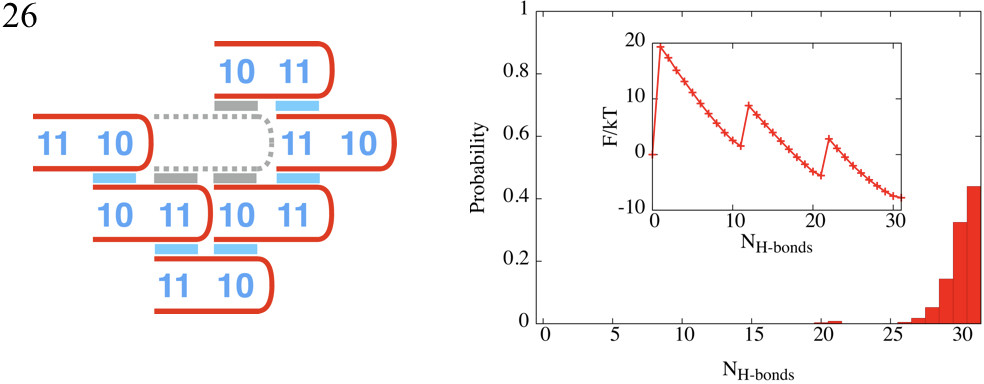}
\end{figure}

\subsection{Local environments with four binding domains}

\begin{figure}[h]
    \centering
    \includegraphics[width=0.90\textwidth]{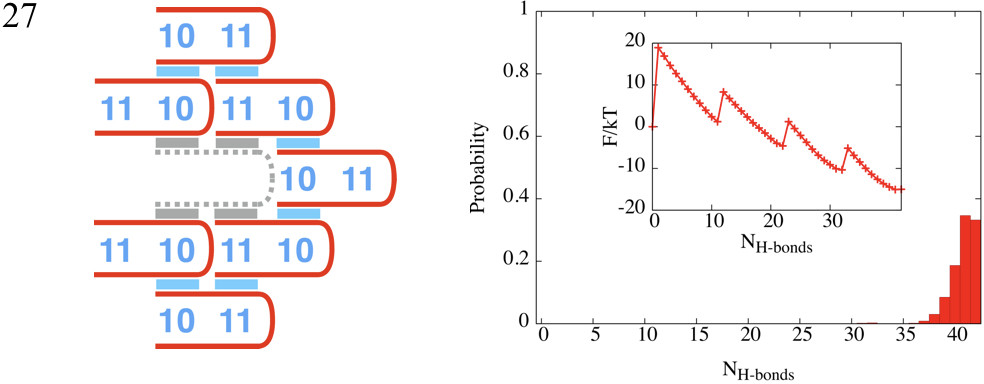}
\end{figure}

\begin{figure}[h]
    \centering
    \includegraphics[width=0.90\textwidth]{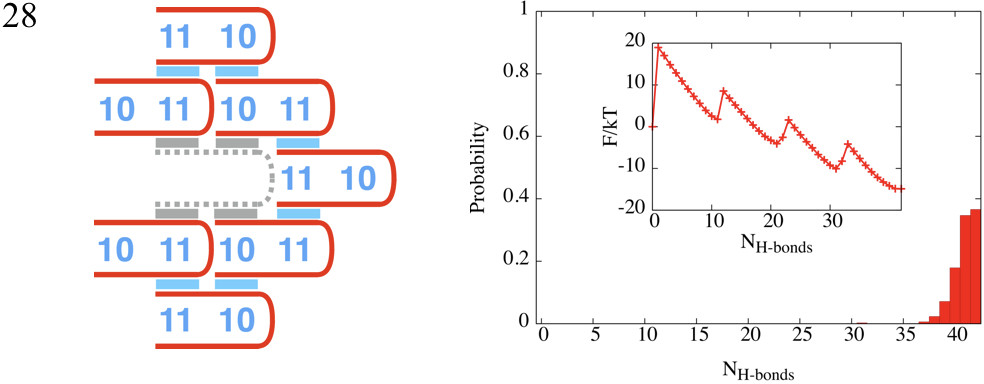}
\end{figure}

\newpage

\begin{figure}[h]
    \centering
    \includegraphics[width=0.90\textwidth]{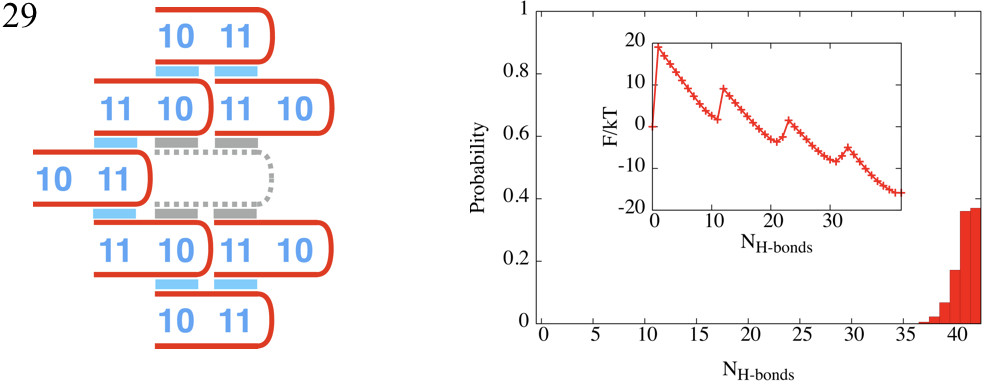}
\end{figure}

\begin{figure}[h]
    \centering
    \includegraphics[width=0.90\textwidth]{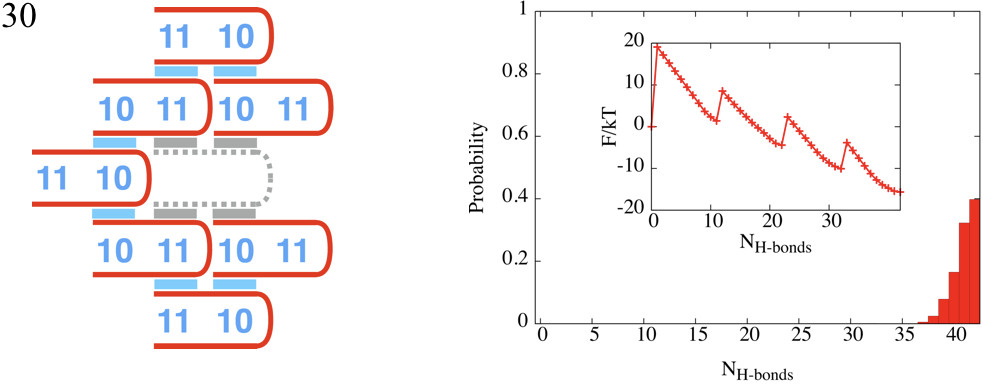}
\end{figure}

\begin{figure}[h]
    \centering
    \includegraphics[width=0.90\textwidth]{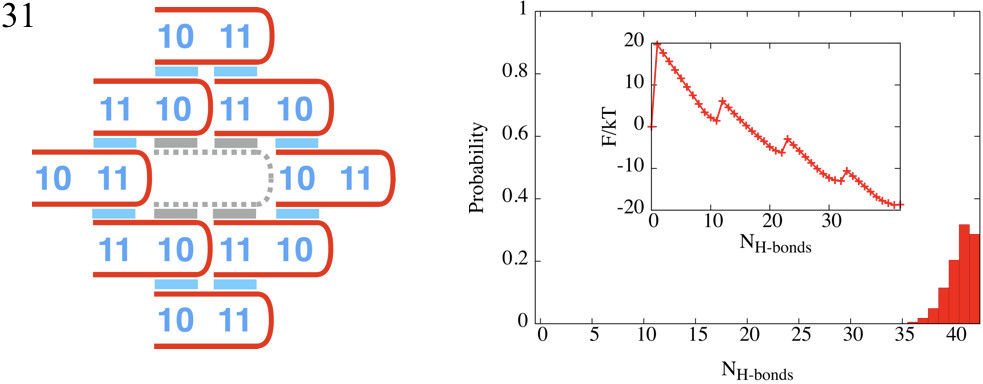}
\end{figure}

\newpage

\begin{figure}[h]
    \centering
    \includegraphics[width=0.90\textwidth]{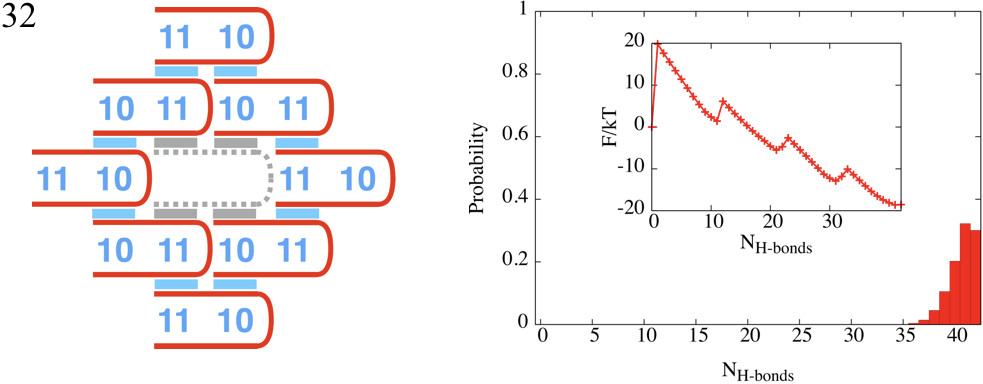}
\end{figure}

\newpage

\section{Long range bridging - example}

\begin{figure}[h]
    \centering
    \includegraphics[width=0.30\textwidth]{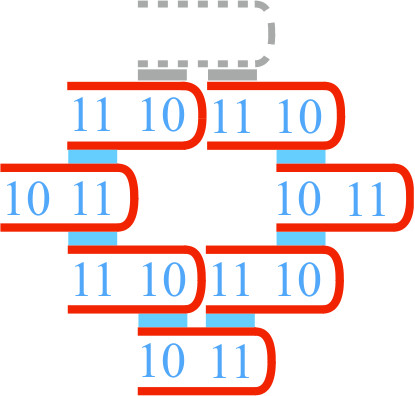}
        \caption{Example of a ``bridging tile'' (coloured grey) whose association involves bringing together two initially distant sections of the assembly fragment.  The binding domains in the assembly fragment that are hybridized are shown in blue. The numbers 10 and 11 indicate the length (in bases) of the binding domains positioned above. }\label{fig:random_lengths}
\end{figure}


\newpage

\section{Deviations from detailed balance I}


\begin{figure}[h]
    \centering
    \includegraphics[width=0.88\textwidth]{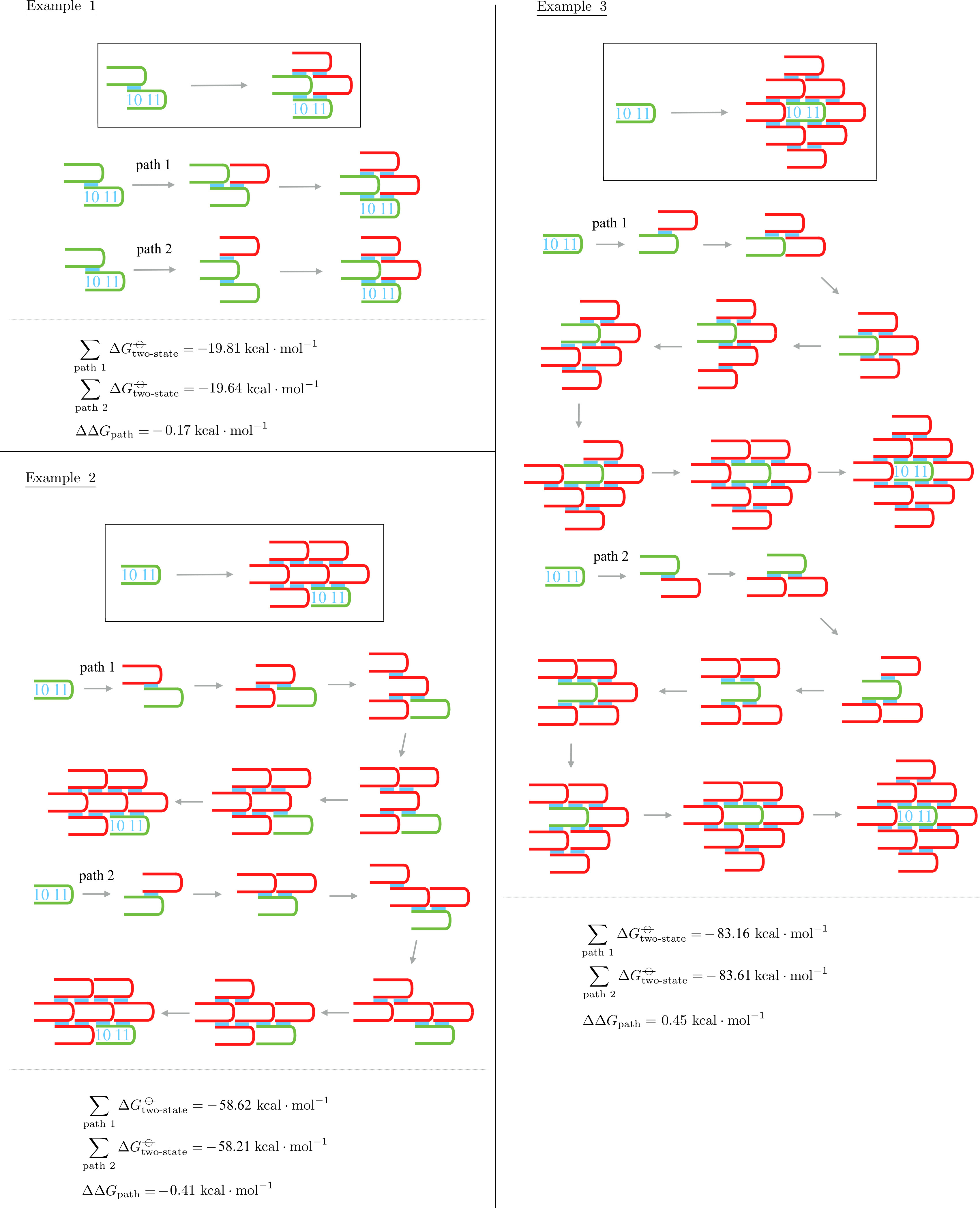}
\caption{Calculation of free-energy differences using different reaction paths. Calculation performed at $T = 50^\circ\text{C}$ using tile-association free energies determined using oxDNA.}\label{fig:path_dg_tests}
\end{figure}

\newpage

\section{Deviations from detailed balance II}

\begin{figure}[h]
    \centering
    \includegraphics[width=0.45\textwidth]{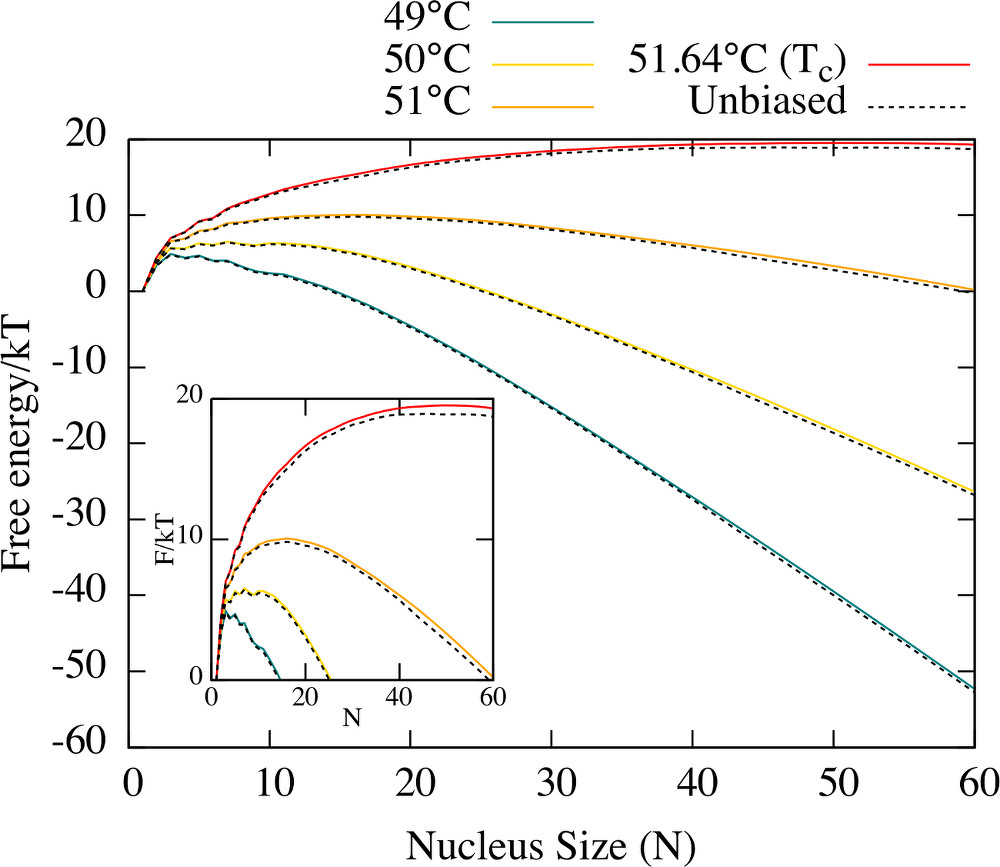}
\caption{Free-energy profiles as function of assembly size determined using biased sampling (coloured curves) and unbiased sampling (dashed curves). To make sampling more efficient, the assembly size coordinate was split into 6 sampling windows:  $N_1\in [1,11]$,  $N_2\in [9,21 ]$, $N_3\in [19,31 ]$, $N_4\in [29,41 ]$, $N_5\in [39,51 ]$ and $N_6\in [49,60 ]$; the resulting statistics were then combined by defining $F(N_1=1)=0$, and fixing $F(N_1=10)=F(N_2=10)$, $F(N_2=20)=F(N_3=20)$, $F(N_3=30)=F(N_4=30)$, $F(N_4=40)=F(N_5=40)$ and $F(N_5=50)=F(N_6=50)$.}\label{fig:path_dg_tests_2}
\end{figure}

\section{Estimation of $k_+^0$ from toehold-mediated strand displacement experiments}

To parameterise the rate constant of tile-association ($k_+^0$) we consider the work of D. Zhang and E. Winfree \cite{winfree_kinetics}, in which toehold exchange (TESD) and toehold-mediated (TMSD) strand displacement processes were characterised experimentally and compared against predictions of simple kinetic models, in which both TESD and TMSD are treated as simple bimolecular processes (with two reactants and two products each). 


In the experiments carried out, a spectrofluorimetry analysis was used that provided a real-time probe for changes in the concentration of one of the final products (labelled ``Y'' in the original work). The experimental data was than used to parameterise the rate constants used in the kinetic models, so that a best fit was achieved between experimental and simulated curves. 

Of especial interest for the present work is the analysis made on the TMSD system, which has the particularity that for long toeholds ($\sim 7$ bases or longer) the forward rate constant in the kinetic model corresponds approximately to the rate constant associated with the hybridisation of the toehold with its complementary domain in an initially diffusing single strand, a process which we assume to be kinetically similar to the hybridisation of the first domain during tile-association. 

We choose the value $6\times 10^6$ for $k_+^0$, which corresponds to the rate constant estimated for the hybridisation of a G-C rich toehold in the TMSD system with a length of 10 bases  (the maximum toehold length considered in Ref. \cite{winfree_kinetics}), with a complementary domain in a diffusing single strand.

\bibstyle{aip}